{}
{}

\documentclass[12pt,a4paper]{article}

\newif\ifpublic\publicfalse
\newif\ifniklas\niklastrue
\newif\ifniklas\niklastrue

\usepackage{ifpdf}
\ifniklas\ifpdf\publictrue\else\publicfalse
\PassOptionsToPackage{hypertex}{hyperref}
\PassOptionsToPackage{draft}{graphicx}
\fi\fi

\usepackage{subfigure}
\usepackage{color}
\usepackage{tikz}
\usetikzlibrary{decorations.markings}
\usetikzlibrary{shapes,arrows,automata,positioning}
\usepackage{tikz-3dplot}

\tikzstyle{decision} = [diamond, draw, fill=blue!20, 
    text width=4.5em, text badly centered, node distance=3cm, inner sep=0pt]
\tikzstyle{block} = [rectangle, draw, fill=blue!20, 
    text width=5em, text centered, rounded corners, minimum height=4em]
\tikzstyle{line} = [draw, -latex']
\tikzstyle{cloud} = [draw, ellipse,fill=red!20, node distance=3cm,
    minimum height=2em]

\usepackage[utf8]{inputenc}
 \usepackage[T1]{fontenc}
\usepackage{amsmath,amssymb,amsthm}
\ifpublic\else\usepackage{showkeys}\fi
\usepackage[bookmarks=true,hyperfigures=true]{hyperref}
\usepackage{graphicx}
\usepackage{dsfont}
\usepackage{xcolor}
\usepackage[nosort]{cite}
\usepackage[bulletsep]{collref}
\usepackage{bm}
\usepackage{multirow}
\usepackage{enumitem}

\usepackage{comment}
\ifniklas

\def\showkeysrefformat#1{{\normalfont\tiny\ttfamily#1}}
\makeatletter
\def\SK@@ref#1>#2\SK@{%
 {\@inlabelfalse\leavevmode\vbox to\z@{%
 \vss\SK@refcolor\rlap{\vrule\raise .75em%
  \hbox{\showkeysrefformat{#2}}}}}}
\makeatother
\fi

\usepackage[a4paper,text={160mm,247mm},centering]{geometry}

\allowdisplaybreaks[3]

\numberwithin{equation}{section}

\usepackage[font=small,labelfont=bf,width=0.85\textwidth]{caption}

\expandafter\def\expandafter\bfseries\expandafter{\bfseries\ifmmode\else\boldmath\fi}
\expandafter\def\expandafter\mdseries\expandafter{\mdseries\ifmmode\else\unboldmath\fi}
\expandafter\def\expandafter\normalfont\expandafter{\normalfont\ifmmode\else\unboldmath\fi}

\RequirePackage{verbatim}

\makeatletter
\newwrite\bibinl@out
{\immediate\closeout\bibinl@out\@esphack}
\makeatother

\usepackage{todonotes}
\newcommand{\AP}[1]{\todo[inline,color=green!20!white]{\footnotesize{\textbf{AP:}} \footnotesize{#1}}}

\usepackage{fixmath}

\newcommand{\alg}[1]{\mathfrak{#1}}

\newcommand{\galg}{\mathfrak{g}}
\newcommand{\kQ}{\mathfrak{Q}}
\newcommand{\kU}{\mathfrak{U}}
\newcommand{\kP}{\mathfrak{P}}
\newcommand{\kV}{\mathfrak{V}}
\newcommand{\kS}{\mathfrak{S}}
\newcommand{\kC}{\mathfrak{C}}
\newcommand{\kH}{\mathfrak{H}}

\newcommand{\lN}{\mathcal{N}}
\newcommand{\lH}{\mathcal{H}}

\newcommand{\lR}{\mathcal{R}}
\newcommand{\lS}{\mathcal{S}}

\newcommand{\lO}{\mathcal{O}}

\newcommand{\bbR}{\mathbb{R}}

\newcommand{\su}{\mathfrak{su}}
\newcommand{\psu}{\mathfrak{psu}}
\newcommand{\adss}{{\rm AdS}}
\newcommand{\rmr}{\mathrm{r}}
\newcommand{\tA}{\text{A}}
\newcommand{\tB}{\text{B}}
\newcommand{\tC}{\text{C}}
\newcommand{\tL}{\text{L}}
\newcommand{\tR}{\text{R}}

\newcommand{\comP}[1]{{\bf\textcolor{magenta} {P:} \textcolor{magenta}{#1}}}

\newcommand{\comA}[1]{{\bf\textcolor{blue} {Ana:} \textcolor{blue}{#1}}}

\RequirePackage{amsmath}
\makeatletter
\newcommand{\brk@ord}{\bBigg@{0}}
\newcommand{\brk@ordl}{\mathopen\brk@ord}
\newcommand{\brk@ordr}{\mathclose\brk@ord}
\newcommand{\brk@ordm}{\mathrel\brk@ord}
\newcommand{\brk@var}{\brk@ord}
\newcommand{\brk@varl}{\left}
\newcommand{\brk@varr}{\right}
\newcommand{\brk@varm}{\mathrel\brk@var}
\newcommand{\brk@altname}[3]{\expandafter\def\csname#2\expandafter\@gobble\string#1\endcsname{#1[#3]}}
\newcommand{\brk@usearg}[3]{%
  \def\brk@star{*}\def\brk@blank{}\def\brk@arg{#1}%
  \ifx\brk@arg\brk@blank\def\brk@arg{brk@ord}\fi%
  \ifx\brk@arg\brk@star\def\brk@arg{brk@var}\fi%
  \csname\brk@arg #2\endcsname#3}

\newcommand{\DeclareMathBrackets}[3]{
  \newcommand{#1}[2][]{\brk@usearg{##1}{l}{#2}##2\brk@usearg{##1}{r}{#3}}
  \brk@altname{#1}{big}{big}\brk@altname{#1}{lr}{*}}
\newcommand{\DeclareMathBiBrackets}[4]{
  \newcommand{#1}[3][]{\brk@usearg{##1}{l}{#2}##2#3##3\brk@usearg{##1}{r}{#4}}
  \brk@altname{#1}{big}{big}\brk@altname{#1}{lr}{*}}
\newcommand{\DeclareMathBiMBracketsStar}[4]{
  \newcommand{#1}[3][]{\brk@usearg{##1}{l}{#2}##2\brk@usearg{##1}{m}{#3}##3\brk@usearg{##1}{r}{#4}}
  \brk@altname{#1}{bi}{big}}
\newcommand{\DeclareMathBiBracketsStar}[4]{
  \newcommand{#1}[3][]{\brk@usearg{##1}{l}{#2}##2\brk@usearg{##1}{}{#3}##3\brk@usearg{##1}{r}{#4}}
  \brk@altname{#1}{big}{big}}

\makeatother

\DeclareMathBrackets{\brk}{(}{)}
\DeclareMathBrackets{\sbrk}{[}{]}
\DeclareMathBrackets{\set}{\{}{\}}
\DeclareMathBrackets{\abs}{|}{|}
\DeclareMathBrackets{\eval}{.}{|}
\DeclareMathBrackets{\spn}{\langle}{\rangle}
\DeclareMathBiBrackets{\comm}{[}{,}{]}
\DeclareMathBiBrackets{\acomm}{\{}{,}{\}}
\DeclareMathBiBrackets{\gcomm}{[}{,}{\}}

\newcommand{\lP}{\mathcal{P}}
\newcommand{\lI}{\mathcal{I}}

\newcommand{\bP}{\mathbb{P}}
\newcommand{\bI}{\mathbb{I}}

\def\[{\begin{equation}}
\def\]{\end{equation}}

\providecommand{\href}[2]{#2}

\makeatletter
\def\mr@ignsp#1 {\ifx\:#1\@empty\else #1\expandafter\mr@ignsp\fi}%
\newcommand{\multiref}[1]{\begingroup
\xdef\mr@no@sparg{\expandafter\mr@ignsp#1 \: }%
\def\mr@comma{}%
\@for\mr@refs:=\mr@no@sparg\do{\mr@comma\def\mr@comma{,}\ref{\mr@refs}}%
\endgroup}
\renewcommand{\eqref}[1]{(\multiref{#1})}
\makeatother

\makeatletter
\newcommand{\namedref}[2]{\hyperref[#2]{#1~\ref*{#2}}}
\newcommand{\secref}{\@ifstar{\namedref{Section}}{\namedref{Sec.}}}
\newcommand{\appref}{\@ifstar{\namedref{Appendix}}{\namedref{App.}}}
\newcommand{\tabref}{\@ifstar{\namedref{Table}}{\namedref{Tab.}}}
\newcommand{\figref}{\@ifstar{\namedref{Figure}}{\namedref{Fig.}}}
\makeatother

\usepackage{mathtools}

\DeclarePairedDelimiterX\braket[2]{\langle}{\rangle}{#1 \delimsize\vert #2}

\providecommand{\hypersetup}[1]{}

\hypersetup{plainpages=false}
\hypersetup{pdfpagemode=UseNone}
\hypersetup{bookmarksnumbered=true}
\hypersetup{pdfstartview=FitH}
\hypersetup{colorlinks=false}
\hypersetup{citebordercolor={.5 1 .5}}
\hypersetup{urlbordercolor={.5 1 1}}
\hypersetup{linkbordercolor={1 .7 .7}}

\makeatletter
\let\@keywords\@empty
\let\@subject\@empty
\providecommand{\keywords}[1]{\gdef\@keywords{#1}}
\providecommand{\subject}[1]{\gdef\@subject{#1}}
\def\thetitle{\@title}
\def\theauthor{\@author}
\def\thesubject{\@subject}
\def\thedate{\@date}
\def\thekeywords{\@keywords}
\makeatother
\AtBeginDocument{
\hypersetup{pdftitle={\thetitle}}%
\hypersetup{pdfauthor={\theauthor}}%
\hypersetup{pdfsubject={\thesubject}}%
\hypersetup{pdfkeywords={\thekeywords}}%
}

\newcommand{\ana}[1]{$\framebox{\tiny a}$\ \textbf{\texttt{{\color{red}\footnotesize#1}}}}
\newcommand{\mdlnote}[1]{$\framebox{\tiny MdL}$\ \textbf{\texttt{{\color{blue}\footnotesize#1}}}}

\title{Integrable deformations of AdS/CFT}
\author{ Marius de Leeuw, Anton Pribytok, Ana L. Retore and Paul Ryan}

\begin{document}

\pdfbookmark[1]{Title Page}{title}
\thispagestyle{empty}


\vspace*{2cm}
\begin{center}%
\begingroup\Large\bfseries\thetitle\par\endgroup
\vspace{1cm}

\begingroup\scshape\theauthor\par\endgroup
\vspace{5mm}%

\begingroup\itshape
School of Mathematics
\& Hamilton Mathematics Institute\\
Trinity College Dublin\\
Dublin, Ireland
\par\endgroup
\vspace{5mm}

\begingroup\ttfamily
$\{$mdeleeuw,
apribytok,
retorea,
pryan$\}$@maths.tcd.ie
\par\endgroup

\vfill

\textbf{Abstract}\vspace{5mm}

\begin{minipage}{12.7cm}
In this paper we study in detail the deformations introduced in \cite{deLeeuw:2020ahe} of the integrable structures of the AdS${}_{2,3}$ integrable models. We do this by embedding the corresponding scattering matrices into the most general solutions of the Yang-Baxter equation. We show that there are several non-trivial embeddings and corresponding deformations. 
We work out crossing symmetry for these models and study their symmetry algebras and representations. In particular, we identify a new elliptic deformation of the AdS$_3\times\mathrm{S}^3\times\mathrm{M}^4$ string sigma model. 
\end{minipage}

\vspace*{4cm}

\end{center}

\newpage


\tableofcontents
\section{Introduction}

The main hallmark of integrable field theories is the factorisation of scattering events into sequences of two-body scattering processes \cite{ZAMOLODCHIKOV1979253,Zamolodchikov:1990bu}. This is due to the presence of a tower of conserved charges, which severely restrict possible scattering events. In particular, scattering is purely elastic and there is no particle production or annihilation. The property that three-body scattering factorises into a sequence of two-body scattering events in a consistent way  imposes the following constraint on the two-body $S$-matrix 
\begin{equation}
\lS_{23}\lS_{12}\lS_{23}=\lS_{12}\lS_{23}\lS_{12}.
\end{equation}
This is the celebrated Yang--Baxter equation
\cite{Jimbo:1989qm,jimbo1990yang,Perk2006yang}. 


Integrable field theories have made numerous appearances in the $ \, $ context $ \, $ of $ \, $ the 
AdS/CFT correspondence \cite{Maldacena:1998zhr,Witten:1998qj,Klebanov:2004ya} and on various string backgrounds the $1+1$ dimensional theory on the string worldsheet defines an integrable QFT, see for example\cite{Fateev:1996ea,Metsaev:1998it,Beisert:2010jr,Arutyunov:2009ga,Sfondrini:2014via}. This is achieved by fixing the uniform light-cone gauge and decompactifying the worldsheet to a plane where the notion of asymptotic states and hence an S-matrix can be defined. From here, one can study scattering in perturbation theory for example.

It is clearly not feasible to explicitly compute a scattering process to all orders in perturbation theory. Instead one can make use of symmetry considerations. While the fixing of the light-cone gauge breaks some of the isometries of the initial background the residual symmetry is often highly constraining. We will focus our attention in this work on the ${\rm AdS}_3 \times S^3\times T^4$, ${\rm AdS}_3 \times S^3\times S^3 \times S^1$ and ${\rm AdS}_2 \times S^2\times T^6$ backgrounds which are known to lead to integrable QFTs, see for example \cite{AdS2_2011,Borsato:2014hja,Hoare:2014kma,Borsato:2015mma}. At one loop the residual symmetry algebras can be determined by explicit calculation. From here one can conjecture the exact all-loop symmetry algebra. 
The symmetry algebras together with integrability (imposing the Yang-Baxter equation) completely determine the $S$-matrix up to the dressing phase, which is constrained by crossing symmetry instead.

An interesting question is the study of deformations of integrable QFTs which preserve integrablity. Numerous such deformations of the models described above have been constructed, for example $\eta$ and $\lambda$ deformations \cite{Delduc:2013qra,Delduc:2014kha,Hollowood:2014qma,Arutyunov:2013ega,Sfetsos:2013wia,Sfetsos:2014cea}. In these cases the isometry algebra is known to undergo a $q$-deformation. For instance, in the case of the $\adss_5\times S^5$ background whose lightcone gauge fixed model has $\su(2|2)_{\rm ce}$ symmetry, the deformed model has $U_q(\psu(2|2)_{\rm ce})$ symmetry \cite{Beisert:2008tw,deLeeuw:2011jr}. These deformations have also been worked out for the AdS$ _3 $ models and various generalisations and other deformations of our mentioned backgrounds have been constructed, see e.g.\cite{Delduc:2014kha,Lukyanov:2012zt,Hoare:2014oua,Delduc:2018xug,Seibold:2019dvf,Bocconcello:2020qkt,Garcia:2021iox,Seibold:2021lju}.

The deformations just discussed are usually constructed at the level of the superstring action and it is not always clear how these deformations lift to the level of the worldsheet $S$-matrix and the corresponding symmetry algebra. In \cite{deLeeuw:2020ahe} we classified all possible $S$-matrices of integrable systems of so-called $6$- and $8$-vertex type. Physically, these correspond to $S$-matrices which preserve fermion number. We can embed the superstring $S$-matrices in these models by fixing various free functions and parameters. By varying these parameters we subsequently obtain integrable deformations of the $S$-matrices. Hence, we actually find a complete classification of possible integrable deformations to the aforementioned superstring $S$-matrices. In this way we therefore identify the possible integrable deformations of these holographic models. 

\begin{figure}[h]
	\hspace{3cm}	\includegraphics[width=10cm]{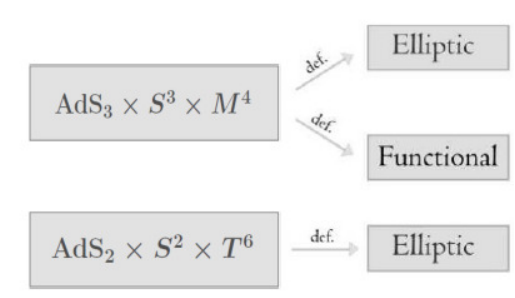}
	\caption{New deformations of AdS$ _3 $ and AdS$ _2 $ $ R $-matrices.}
	\label{deformationsdiagram}
\end{figure}

We find an interesting possible deformation structure, summarised in Figure \ref{deformationsdiagram}. The massive scattering matrix of the AdS$_2$ superstring only admits one integrable deformation. It is an elliptic deformation, parameterized by the elliptic modulus. It exhibits the same Lie algebra, but in a different representation. Remarkably, the deformation does seem to affect the higher Yangian symmetry generators and hence might define a new type of quantum algebra.

More interestingly the $S$-matrix of the $ {{\rm AdS}}_3\times \mathrm{S}^3\times \mathrm{M}^4$ string sigma model admits \textit{two} distinct deformations. The first deformation of the $S$-matrix is related to the quantum deformation found in \cite{Hoare:2014oua,Seibold:2021lju} and the only extra degrees of freedom stem from the fact our functions are unconstrained and do not need to satisfy relations involving the physical constants such as the mass. This is why we refer to such deformations as \textit{functional} deformations.

The second deformation is an elliptic deformation and it is a novel deformation. We show that it satisfies all the usual physical requirements such as crossing symmetry and braiding unitarity. Remarkably, for this model part of the symmetry algebra seems to be broken for non-zero values of the deformation parameter. It would be very interesting to see if the corresponding sigma model can be found. In particular, we find that the elliptic deformation is not a further deformation of the q-deformed model\cite{Hoare:2014oua}.

Finally, it is also interesting to notice that all the deformations of AdS$ _{2,3} $ $S$-matrices satisfy the free fermion condition\cite{deLeeuw:2020bgo}. This allows for a great simplification in known results for lower-dimensional instances of AdS/CFT, including backgrounds supported by various fluxes. 
For 6-vertex like AdS$ _3 $ models and its deformations, the transfer matrix was rewritten for an arbitrary number of sites in a free-fermion form using Bogoliubov transformations.

\paragraph{Outline of this paper} 
In this paper we examine the structure of these integrable deformations in more detail. In Section \ref{rmatrices} we review the $6$- and $8$-vertex $R$-matrices which the conventional AdS${}_{2,3}$ integrable systems can be embedded in. In Section \ref{AdS2def} we discuss the AdS$_2$ deformations while in Section \ref{deformations} we discuss the AdS$ _3 $ deformations and extend our deformed $R$-matrices for same-chirality scattering in order to account for scattering processes between particles with opposite chirality. In Section \ref{Comparison} we give the explicit map to recover known $ {{\rm AdS}}_3$ $S$-matrices, as well as its q-deformations. The deformed symmetry algebras are presented in Section \ref{symmetryalgs}. Finally, in Section  \ref{crossing} we discuss crossing symmetry and  provide the crossing equations for all the deformations explicitly. We end with discussion and conclusions.

\section{Review and notation}\label{rmatrices}

In order to preserve fermion number, for boson $\phi$  and fermion $ \psi $ the allowed scattering processes are 
\begin{align*}
\phi\phi\rightarrow \phi \phi+\psi \psi \\
\psi\psi\rightarrow \phi \phi+\psi \psi \\
\phi\psi\rightarrow \phi \psi+\psi \phi \\
\psi\phi\rightarrow \psi \phi+\phi \psi
\end{align*}
with some weighting associated to each process, or in matrix form we have
\begin{align}
\begin{split}
R(u,v)  =\begin{pmatrix}
r_1 & 0 & 0 & r_8 \\
0 & r_2 & r_6 & 0 \\
0 & r_5 & r_3 & 0 \\
r_7 & 0 & 0 & r_4 \\
\end{pmatrix}
\end{split}
\label{generalR8vertex}
\end{align}
where $u,v$ are the spectral parameters and each of the functions $r_j$ is assumed to depend on them. We will also use the $(p,q)$ instead of $(u,v)$ throughout the text to emphasize that these parameters correspond to the momenta of the particles being scattered. The $R$-matrices classified in \cite{deLeeuw:2020ahe} are the most general $R$-matrices of the above form.

The $R$-matrices of this type can be separated into two categories -- so-called $6$-vertex and $8$-vertex models. Physically, $6$-vertex models are those for which spin is conserved in scattering processes and so a boson pair cannot scatter to produce a fermion pair and vice-versa and as a result $r_7=r_8=0$. In \cite{deLeeuw:2020ahe} both $6$- and $8$-vertex were further divided into two subcategories, dubbed $A$ and $B$. These categories are described by the free fermion condition \cite{deLeeuw:2020bgo}. For applications to AdS integrable systems it is only the $6$-vertex B and $8$-vertex B which are relevant. Hence, for notation simplicity we will simply refer to them as $6$-vertex or 6vB and $8$-vertex or 8vB.

\subsubsection*{$6$-vertex}

The $R$-matrix for the 6-vertex case has $r_7=r_8=0$ and can be written as
\begin{align}
& r_1(p,q)=\frac{h_2(q)-h_1(p)}{h_2(p)-h_1(p)},\nonumber\\
& r_2(p,q) =(h_2(p)-h_2(q))X(p)Y(p),\nonumber\\
& r_3(p,q) =\frac{h_1(p)-h_1(q)}{(h_2(p)-h_1(p))(h_2(q)-h_1(q))}\frac{1}{X(q)Y(q)},\nonumber\\
& r_4(p,q) =\frac{h_2(p)-h_1(q)}{h_2(q)-h_1(q)}\frac{X(p)Y(p)}{X(q)Y(q)},\nonumber\\
& r_5(p,q) =\frac{Y(p)}{Y(q)},\nonumber\\
& r_6(p,q) =\frac{X(p)}{X(q)}.
\label{6vertex}
\end{align}
$h_1,\, h_2,\, X,\, Y$ are free functions. The Yang-Baxter equation is satisfied for any choice of them. 
Compared to \cite{deLeeuw:2020ahe} we have some extra functions in our $R$-matrix, namely $ X $ and $ Y $. This is due to the fact that the $R$-matrix in \cite{deLeeuw:2020ahe}  was solved by fixing some functions in the Hamiltonian using identifications such as normalization and local basis transformations. The new functions $X,Y$ simply correspond to a twist and are important for the identification with the AdS$_3$ models and in order to obtain the crossing equations presented in Section \ref{crossing}.

\subsubsection*{$8$-vertex}

The $R$-matrix is most conveniently written using Jacobi elliptic functions and we use the shorthand notation 
\begin{align}
&\mathrm{sn} =\mathrm{sn}(u-v,k^2) ,
&&\mathrm{cn} =\mathrm{cn}(u-v,k^2) ,
&&\mathrm{dn} =\mathrm{dn}(u-v,k^2) ,
\end{align}
to denote the elliptic functions of modulus $k$. To avoid potential ambiguities in conventions let us stress that we follow the convention that the elliptic functions written above is how they would be entered in Mathematica, and so we have, for example 
\begin{equation}
\mathrm{dn}^2+k^2\, \mathrm{sn}^2=1\,.
\end{equation}
The entries of the $R$-matrix are given by
\begin{align}
r_1 &= 
\frac{1}{\sqrt{\sin \eta(u)}\sqrt{\sin\eta(v)}}  \bigg[\sin\eta_+\frac{\mathrm{cn}}{\mathrm{dn}} 
-\cos\eta_+ \mathrm{sn}\bigg], \nonumber\\
r_2 &= 
\frac{-1}{\sqrt{\sin \eta(u)}\sqrt{\sin\eta(v)}}  \bigg[\cos\eta_-\mathrm{sn} +\sin\eta_-\frac{\mathrm{cn}}{\mathrm{dn}} 
\bigg],\nonumber\\
r_3 &= 
\frac{-1}{\sqrt{\sin \eta(u)}\sqrt{\sin\eta(v)}}  \bigg[\cos\eta_-\mathrm{sn}  - \sin\eta_-\frac{\mathrm{cn}}{\mathrm{dn}} 
\bigg],\nonumber\\
r_4 &= 
\frac{1}{\sqrt{\sin \eta(u)}\sqrt{\sin\eta(v)}}  \bigg[\sin\eta_+\frac{\mathrm{cn}}{\mathrm{dn}} 
+\cos\eta_+ \mathrm{sn}\bigg],\nonumber\\
r_5&=r_6=1,\nonumber\\
r_7 &= r_8 = k\, \mathrm{sn}\frac{\mathrm{cn}}{\mathrm{dn}},
\label{8vertex}
\end{align}
where $\eta_\pm = \frac{\eta(u) \pm \eta(v)}{2}$.

\section{Deforming the ${\rm AdS}_{2}$ $S$-matrix}\label{AdS2def}

The massive $S$-matrix for the $\mathrm{AdS}_{2}$ integrable model \cite{Hoare:2014kma} can only be embedded in the model 8vB due to the presence of the components $r_{7,8}$. In order to construct the embedding we need to first perform various integrability-preserving transformations on the $R$-matrix which can be found in \cite{deLeeuw:2020xrw}.

The main issue to be overcome is that the spectral parameters appearing in both models are different despite being denoted by the same letters $u$ and $v$. To get around this we need to transform $(u,v) \mapsto (G(u),G(v))$ in one of the $R$-matrices and we take this to be in $R^{8vB}$. Notice, also that the R-matrix as in equation \eqref{8vertex} is the boson-boson one, so in order to compare it with \cite{Hoare:2014kma} one needs to do the appropriate modifications.

In \cite{deLeeuw:2020ahe} we compared the corresponding Hamiltonians to show that $R^{{\rm AdS}_2}$ is a special case of $R^{8vB}$, but it is also instructive to compare the $R$-matrices directly.
We start by considering the $(1,4)$ component of both $R$-matrices which are \cite{Hoare:2014kma}, for $R^{8vB}$ and $R^{{\rm AdS}_2}$ respectively, 
\begin{align}
&(R^{8vB})_{14} = k\, \mathrm{sn}(G(u)-G(v))\frac{\mathrm{cn}(G(u)-G(v))}{\mathrm{dn}(G(u)-G(v))}, \\
&(R^{{\rm AdS}_2})_{14} =  \frac{1}{\sqrt{x^+_u x^-_u x^+_v x^-_v}}\, \frac{\left(x^-_u - \frac{1}{x^+_u}\right)\sqrt{\frac{x^+_u}{x^-_u}}-\left(x^-_v - \frac{1}{x^+_v}\right)\sqrt{\frac{x^+_v}{x^-_v}}}{1-\frac{1}{x^+_u x^-_u x^+_v x^-_v}}
\end{align}
where $x^\pm$ are the Zhukovski variables. Clearly, the $(1,4)$ component of $R^{\rm 8vB}$ is of difference form, that is it only depends on the difference $G(u)-G(v)$ of the spectral parameters. 

Let us expand the $(1,4)$ component of the AdS$ _2 $ $ R $-matrix in $u$ around $v$. We find 
\begin{equation}
\frac{\left(x^+ x^-\right)'}{2\sqrt{x^-}\sqrt{x^+}(x^+x^--1)}(u-v)+\mathcal{O}\left((u-v)^2 \right).
\end{equation}
In order to be purely of difference form we must have that the coefficient of $u-v$ is a constant which we denote $A$:
\begin{equation}
\frac{\left(x^+ x^-\right)'}{2\sqrt{x^-}\sqrt{x^+}(x^+x^--1)} = A.
\end{equation}
Hence, after reinstating the $G$ dependence, we solve to obtain
\begin{equation}\label{eq:AdS2Xpm}
x^+(v) = \frac{{\rm Tanh}\left(A G(v) +\frac{c_1}{2}\right)}{x^-(v)}\,.
\end{equation}
This completely fixes $G$ in terms of $x^\pm$.

After substituting \eqref{eq:AdS2Xpm} back into the $(1,4)$ component of the AdS$ _2 $ $ R $-matrix we find that it reduces to simply\footnote{Working in an appropriate region such that we avoid branch cut issues.}
\begin{equation}
(R^{{\rm AdS}_2})_{14} = -{\rm Tanh}\left(A(G(u)-G(v))\right)\,.
\end{equation}
A comparison with the $(1,4)$ component of the 8vB $ R $-matrix then tells us that we should take the limit $k\rightarrow \infty$ in order to have this entry reduce to ${\rm Tanh}$ and furthermore the precise agreement requires that $A=-i$ and we can take $c_1=0$, and so we find that \footnote{This relation is somewhat reminiscent of the procedure to arrive at the massless gamma variable \cite{Fontanella:2019baq,Fontanella:2019ury}.}
\begin{equation}
x^+(u) = -\frac{{\rm Tan}^2(G(u))}{x^-(u)}\,.
\end{equation}

\medskip

Next, we make the substitution $\eta(u) \rightarrow {\rm arccot}\left(k F(u)\right)$ and expand the 8vB $ R $-matrix around $k \rightarrow \infty$. By subsequently expanding around $u=v$ we find that setting
\begin{equation}
F(u) = -\frac{1}{2}{\rm csc}(G(u))\, {\rm sec}(G(u))\, \frac{{\rm cot}(G(u))\,x^-+i}{{\rm cot}(G(u))\,x^--i}
\end{equation}
indeed reproduces the AdS${}_2$ $R$-matrix. Notice that $x^-$ is in principle defined implicitly via the shortening condition
\begin{equation}
x^++\frac{1}{x^+}-x^- - \frac{1}{x^-} = \frac{2im}{h},
\end{equation}
where $ m $ is the mass, and the $ h $ is the coupling constant. But this relation is not needed for the mapping between the two models. Reversely, we can define $m$ in terms of $F,\,G$ in this way.

\paragraph{Deformation}

In order to embed $R_{{\rm AdS}_2}$ into $R^{\rm 8vB}$ we sent the elliptic modulus $k$ to infinity. We then immediately see a source of deformation comes from flowing away from infinity to finite values of $k$. This is the only available non-trivial source of deformation.

There are also functional deformations corresponding to shifts of the form 
\begin{equation}
\eta(u) \rightarrow \eta(u) + \lambda\, \tilde{\eta}(u)
\end{equation}
where $\tilde{\eta}(u)$ is an arbitrary function and $\lambda$ is a deformation parameter. These deformations correspond to making the mass depend on the spectral parameter $m \mapsto m(u)$.

This completes the embedding of $R_{{\rm AdS}_2}$ into $R^{\rm 8vB}$. Later, we will discuss the deformed symmetry algebra in Section \ref{symmetryalgs}.

\section{Deforming the ${\rm AdS}_{3}$ $S$-matrix}\label{deformations}

 In \cite{deLeeuw:2020ahe,deLeeuw:2020xrw} we derived the most general $4\times4$ integrable $R$-matrices of 8-vertex type. These are exactly the $R$-matrices that are compatible with the graded structure that arises for the holographic integrable models. Hence, by embedding the $R$-matrices of  $\mathrm{AdS}_{3}$ in these $R$-matrices, we find the most general way in which they can be deformed compatible with the splitting in chiral blocks.
 
 In this section we show that the massive $S$-matrix for the AdS$_3$ \cite{Sfondrini:2014via,Borsato:2014exa,Borsato:2015mma} integrable model can be embedded in both $ 6 $-vertex B  and $ 8 $-vertex B extended models. 

Let us explain how to construct the full $16\times 16$ $ R $-matrix given we have the $ 4\times4 $ regular $ R $-matrix, and then we apply this procedure to lift the models described in \eqref{generalR8vertex}-\eqref{8vertex}. 

\subsection{General procedure}\label{subsec:generalprocedure}

 
Following the construction in \cite{Sfondrini:2014via}, we decompose the full $ R $-matrix $\bbR$ as
\begin{equation}
\setcounter{MaxMatrixCols}{16}
{\tiny \bbR=\begin{pmatrix}
	\rmr_1^{{\rm LL}} & 0 & 0 & 0 & 0 & \rmr_8^{{\rm LL}} & 0 & 0 & 0 & 0 & 0 & 0 & 0 & 0 & 0 & 0\\
	0 & \rmr_2^{{\rm LL}} & 0 & 0 & \rmr_6^{{\rm LL}} & 0 & 0 & 0 & 0 & 0 & 0 & 0 & 0 & 0 & 0 & 0\\
	0 & 0 & \rmr_1^{{\rm LR}} & 0 & 0 & 0 & 0 & \rmr_8^{{\rm LR}} & 0 & 0 & 0 & 0 & 0 & 0 & 0 & 0\\
	0 & 0 & 0 & \rmr_2^{{\rm LR}} & 0 & 0 & \rmr_6^{{\rm LR}} & 0 & 0 & 0 & 0 & 0 & 0 & 0 & 0 & 0\\
	0 & \rmr_5^{{\rm LL}} & 0 & 0 & \rmr_3^{{\rm LL}} & 0 & 0 & 0 & 0 & 0 & 0 & 0 & 0 & 0 & 0 & 0\\
	\rmr_7^{{\rm LL}} & 0 & 0 & 0 & 0 & \rmr_4^{{\rm LL}} & 0 & 0 & 0 & 0 & 0 & 0 & 0 & 0 & 0 & 0\\
	0 & 0 & 0 & \rmr_5^{{\rm LR}} & 0 & 0 & \rmr_3^{{\rm LR}} & 0 & 0 & 0 & 0 & 0 & 0 & 0 & 0 & 0\\
	0 & 0 & \rmr_7^{{\rm LR}} & 0 & 0 & 0 & 0 & \rmr_4^{{\rm LR}} & 0 & 0 & 0 & 0 & 0 & 0 & 0 & 0\\
	0 & 0 & 0 & 0 & 0 & 0 & 0 & 0 & \rmr_1^{{\rm RL}} & 0 & 0 & 0 & 0 & \rmr_8^{{\rm RL}} & 0 & 0\\
	0 & 0 & 0 & 0 & 0 & 0 & 0 & 0 & 0 & \rmr_2^{{\rm RL}} & 0 & 0 & \rmr_6^{{\rm RL}} & 0 & 0 & 0\\
	0 & 0 & 0 & 0 & 0 & 0 & 0 & 0 & 0 & 0 & \rmr_1^{{\rm RR}} & 0 & 0 & 0 & 0 & \rmr_8^{{\rm RR}}\\
	0 & 0 & 0 & 0 & 0 & 0 & 0 & 0 & 0 & 0 & 0 & \rmr_2^{{\rm RR}} & 0 & 0 & \rmr_6^{{\rm RR}} & 0\\
	0 & 0 & 0 & 0 & 0 & 0 & 0 & 0 & 0 & \rmr_5^{{\rm RL}} & 0 & 0 & \rmr_3^{{\rm RL}} & 0 & 0 & 0\\
	0 & 0 & 0 & 0 & 0 & 0 & 0 & 0 & \rmr_7^{{\rm RL}} & 0 & 0 & 0 & 0 & \rmr_4^{{\rm RL}} & 0 & 0\\
	0 & 0 & 0 & 0 & 0 & 0 & 0 & 0 & 0 & 0 & 0 & \rmr_5^{{\rm RR}} & 0 & 0 & \rmr_3^{{\rm RR}} & 0\\
	0 & 0 & 0 & 0 & 0 & 0 & 0 & 0 & 0 & 0 & \rmr_7^{{\rm RR}} & 0 & 0 & 0 & 0 & \rmr_4^{{\rm RR}}\\
	\end{pmatrix}} 
\label{fullRmatrixgeneral}
\end{equation}
where $ \bbR\equiv \bbR(u,v) $ and $ \rmr_i^{{\rm AB}}\equiv \rmr_i^{{\rm AB}}(u,v)$ and $ \tA,\tB\in\{\tL,\tR\} $.  The full $ R $-matrix $ \bbR $ satisfies the usual Yang-Baxter equation
\begin{equation}
\bbR_{12}(u,v)\bbR_{13}(u,w)\bbR_{23}(v,w)=\bbR_{23}(v,w)\bbR_{13}(u,w)\bbR_{12}(u,v)
\label{YBE16x16}
\end{equation}
and $ \bbR_{ij} $ act on three vector spaces $ \mathbb{V} $ of dimension four $ \mathbb{V}\otimes \mathbb{V}\otimes\mathbb{V} $.

Let us allow for arbitrary scalar factors $\sigma^{LR}$, then the functions $ \rmr_i^{{\rm AB}}(u,v) $ are matrix elements of the $ 4\times 4 $ $ R $-matrices 
\begin{equation}
\lR^{{\rm AB}}(u,v)=
\sigma^{{\rm AB}}\begin{pmatrix}
r_1^{{\rm AB}} & 0 & 0 & r_8^{{\rm AB}}  \\
0 & r_2^{{\rm AB}} & r_6^{{\rm AB}} & 0  \\
0 & r_5^{{\rm AB}} & r_3^{{\rm AB}} & 0  \\
r_7^{{\rm AB}} & 0 & 0 & r_4^{{\rm AB}}  
\label{4x4Smatrix}
\end{pmatrix}
\end{equation}
where $ \lR^{{\rm LL}} $ and $ \lR^{{\rm RR}} $ are regular, i.e.,
\begin{equation}
\lR^{{\rm LL}}(u,u) \sim \lP\quad \text{and} \quad \lR^{{\rm RR}}(u,u) \sim \lP,
\label{blockregularity}
\end{equation}
and $ \lP $ is the permutation operator  for a Hilbert space of dimension two. The functions $ \sigma^{{\rm AB}}(u,v) $ are at this point arbitrary, but they will have to satisfy certain properties in order for both the blocks and the full $ R $-matrix to satisfy crossing symmetry and braiding unitarity.

The fact that YBE \eqref{YBE16x16} for the full $ 16\times 16 $ is satisfied is equivalent to the fact that the blocks $ \{\lR^{{\rm LL}}(u,v),\,\lR^{{\rm RL}}(u,v),\,\lR^{{\rm LR}}(u,v),\,\lR^{{\rm RR}}(u,v)\} $ satisfy eight Yang-Baxter equations given by all possible ways to distribute two chiralities into three Hilbert spaces:
\begin{equation}
\text{YBE}(\tA,\tB,\tC)= \lR_{12}^{{\rm AB}}(u,v)\lR_{13}^{{\rm AC}}(u,w)\lR_{23}^{{\rm BC}}(v,w)- \lR_{23}^{{\rm BC}}(v,w)\lR_{13}^{{\rm AC}}(u,w)\lR_{12}^{{\rm AB}}(u,v)=0.
\label{YBE8eqs}
\end{equation}
Each $ \lR_{ij}^{{\rm AB}}(u,v) $ acts on $ V^{(\tA)}\otimes V^{(\tB)}\otimes V^{(\tC)} $, where each $ V^{(\tA)} $ is a vector space of dimension two and $ \tA,\tB,\tC\in \{\tL,\tR\} $. So, each two dimensional vector space has a chirality associated to itself.

For example, for $ \tA=\tL,\, \tB=\tR $ and $ \tC=\tL $ we have
\begin{equation}
\text{YBE}(\tL,\tR,\tL)= \lR_{12}^{{\rm LR}}(u,v)\lR_{13}^{{\rm LL}}(u,w)\lR_{23}^{{\rm RL}}(v,w)- \lR_{23}^{{\rm RL}}(v,w)\lR_{13}^{{\rm LL}}(u,w)\lR_{12}^{{\rm LR}}(u,v)=0.
\label{YBELRL}
\end{equation}

So, in other words, if $ \lR^{{\rm RR}},\,\lR^{{\rm LL}},\,\lR^{{\rm RL}}$ and $\lR^{{\rm LR}} $ satisfy all the eight YBE's defined in \eqref{YBE8eqs} then the full $ R $-matrix \eqref{fullRmatrixgeneral} will satisfy the equation \eqref{YBE16x16}.

At this point, it is important to remember that the method developed in \cite{deLeeuw:2020ahe} only allows one to compute regular $ R $-matrices. So, our starting point is to put those matrices in the left-left and right-right blocks and use the six remaining YBE in \eqref{YBE8eqs} to fix the left-right and right-left blocks.

The complete procedure is the following:

\

1. We start by assuming that $ \lR^{{\rm LL}}$ and $ \lR^{{\rm RR}}$ are given by a regular $ 4\times 4 $ $ R $-matrix as for example in equation \eqref{generalR8vertex}. We will assume that both of these blocks can be deformed independently and so we use different names for the functions in the LL and RR blocks. 

\

\noindent Notice that with this YBE$ (\tL,\tL,\tL) $ and YBE$(\tR,\tR,\tR)$ are already satisfied.

\

2. We assume $ \lR^{{\rm LR}}(u,v) $ and $ \lR^{{\rm RL}}(u,v) $ of the form \eqref{4x4Smatrix}. Then, we substitute $ \lR^{{\rm LR}}(u,v) $ together with $ \lR^{{\rm LL}}(u,v) $ in YBE$ (\tL,\tL,\tR) $ and solve them for $ \rmr_i^{{\rm LR}} $. Similarly,  solving in YBE$ (\tR,\tR,\tL) $ we construct $ \rmr_i^{{\rm RL}} $.

\

3. The previous step fixes $ \lR^{{\rm LR}}(u,v) $ and $ \lR^{{\rm RL}}(u,v) $ apart from functions of one variable. We use the remaining four YBE$ (\tA,\tB,\tC) $ to fix these functions.

\

4. The last step is to substitute all $ \rmr_i^{\text{A\,B}} $ in the full $ R $-matrix \eqref{fullRmatrixgeneral} and check that it indeed satisfies YBE \eqref{YBE16x16}. 
 
\

For the final result, one of course needs to check braiding unitarity and to find the crossing relations. Braiding unitarity is checked in the next two sections, while crossing symmetry is discussed on a case by case basis in Section \ref{crossing}.

\subsection{A deformation of AdS$_3 $: 6-vertex B}\label{subsec:full6vB}

Now, for the 6 vertex B model, we present the four blocks obtained by applying the procedure above. 
For the left-left sector, ($ \tA=\tL $ and $ \tB=\tL $) we have
\begin{align}
& r_1^{{\rm LL}}=\frac{h_2^{{\rm L}}(q)-h_1^{{\rm L}}(p)}{h_2^{{\rm L}}(p)-h_1^{{\rm L}}(p)},\quad r_7^{{\rm LL}}=0=r_8^{{\rm LL}},\\
& r_2^{{\rm LL}}=(h_2^{{\rm L}}(p)-h_2^{{\rm L}}(q))X^{{\rm L}}(p)Y^{{\rm L}}(p),\\
& r_3^{{\rm LL}}=\frac{h_1^{{\rm L}}(p)-h_1^{{\rm L}}(q)}{\left(h_2^{{\rm L}}(p)-h_1^{{\rm L}}(p)\right)\left(h_2^{{\rm L}}(q)-h_1^{{\rm L}}(q)\right)}\frac{1}{X^{{\rm L}}(q)Y^{{\rm L}}(q)},\\
& r_4^{{\rm LL}}=\frac{h_2^{{\rm L}}(p)-h_1^{{\rm L}}(q)}{h_2^{{\rm L}}(q)-h_1^{{\rm L}}(q)}\frac{X^{{\rm L}}(p)Y^{{\rm L}}(p)}{X^{{\rm L}}(q)Y^{{\rm L}}(q)},\\
& r_5^{{\rm LL}}=\frac{Y^{{\rm L}}(p)}{Y^{{\rm L}}(q)},\quad r_6^{{\rm LL}}=\frac{X^{{\rm L}}(p)}{X^{{\rm L}}(q)}.
\end{align}
This block satisfies YBE$ (\tL,\tL,\tL)=0 $. Notice that $ \lR^{{\rm LL}} $ looks slightly different from the $ R $-matrix introduced in \cite{deLeeuw:2020ahe}. This is due to a twist\footnote{A twist in the $ R $-matrix is allowed because it keeps YBE invariant. It is also necessary in order to 6-vertex B $ R $-matrix be a deformation of  the $ R $-matrices in \cite{Sfondrini:2014via,Borsato:2014exa,Borsato:2015mma,Hoare:2014oua}.} performed in order to make it satisfy crossing symmetry and compare it with AdS$ _3 $ $ R $-matrix.

For the right-right sector ($ \tA=\tR $ and $ \tB=\tR $) we have
\begin{align}
& r_1^{{\rm RR}}=\frac{h_2^{{\rm R}}(q)-h_1^{{\rm R}}(p)}{h_2^{{\rm R}}(p)-h_1^{{\rm R}}(p)},\quad r_7^{{\rm RR}}=0=r_8^{{\rm RR}}\\
& r_2^{{\rm RR}}=\frac{h_2^{{\rm R}}(p)-h_2^{{\rm R}}(q)}{h_2^{{\rm R}}(p)^2}X^{{\rm R}}(p)Y^{{\rm R}}(p),\\
& r_3^{{\rm RR}}=\frac{h_2^{{\rm R}}(q)^2\left(h_1^{{\rm R}}(p)-h_1^{{\rm R}}(q)\right)}{\left(h_2^{{\rm R}}(p)-h_1^{{\rm R}}(p)\right)\left(h_2^{{\rm R}}(q)-h_1^{{\rm R}}(q)\right)}\frac{1}{X^{{\rm R}}(q)Y^{{\rm R}}(q)},\\
& r_4^{{\rm RR}}=\frac{h_2^{{\rm R}}(q)^2}{h_2^{{\rm R}}(p)^2}\frac{h_2^{{\rm R}}(p)-h_1^{{\rm R}}(q)}{h_2^{{\rm R}}(q)-h_1^{{\rm R}}(q)}\frac{X^{{\rm R}}(p)Y^{{\rm R}}(p)}{X^{{\rm R}}(q)Y^{{\rm R}}(q)},\\
& r_5^{{\rm RR}}=\frac{h_2^{{\rm R}}(q)}{h_2^{{\rm R}}(p)}\frac{X^{{\rm R}}(p)}{X^{{\rm R}}(q)},\quad r_6^{{\rm RR}}=\frac{h_2^{{\rm R}}(q)}{h_2^{{\rm R}}(p)}\frac{Y^{{\rm R}}(p)}{Y^{{\rm R}}(q)}.
\end{align}
This block is independent of $ \lR^{{\rm LL}} $ block and from the $ R $-matrix introduced in \cite{deLeeuw:2020ahe} again due to a twist. Also, $ \lR^{{\rm RR}} $ satisfies YBE$ (\tR,\tR,\tR)=0$.
So, the functions in $ \lR^{{\rm LL}} $ and $ \lR^{{\rm RR}} $ are independent and what connects them are the blocks of opposite chirality introduced now.

The right-left sector ($ \tA=\tR $ and $ \tB=\tL $) is given by
 \begin{align}
 & r_1^{{\rm RL}}=1,\quad r_5^{{\rm RL}}=0=r_6^{{\rm RL}}\\
 & r_2^{{\rm RL}}=-\frac{h_2^{{\rm R}}(p)^2}{h_2^{{\rm R}}(p)-h_1^{{\rm R}}(p)}\frac{1+h_1^{{\rm L}}(q)h_1^{{\rm R}}(p)}{1+h_1^{{\rm L}}(q)h_2^{{\rm R}}(p)}\frac{1}{X^{{\rm R}}(p)Y^{{\rm R}}(p)},\\
 & r_3^{{\rm RL}}=-\left(h_2^{{\rm L}}(q)-h_1^{{\rm L}}(q)\right)\frac{1+h_2^{{\rm L}}(q)h_2^{{\rm R}}(p)}{1+h_1^{{\rm L}}(q)h_2^{{\rm R}}(p)}X^{{\rm L}}(q)Y^{{\rm L}}(q),\\
 & r_4^{{\rm RL}}=-h_2^{{\rm R}}(p)^2\frac{h_2^{{\rm L}}(q)-h_1^{{\rm L}}(q)}{h_2^{{\rm R}}(p)-h_1^{{\rm R}}(p)}\frac{1+h_2^{{\rm L}}(q)h_1^{{\rm R}}(p)}{1+h_1^{{\rm L}}(q)h_2^{{\rm R}}(p))}\frac{X^{{\rm L}}(q)Y^{{\rm L}}(q)}{X^{{\rm R}}(p)Y^{{\rm R}}(p)},\\
 & r_7^{{\rm RL}}=i\,h_2^{{\rm R}}(p)\frac{h_2^{{\rm L}}(q)-h_1^{{\rm L}}(q)}{1+h_1^{{\rm L}}(q)h_2^{{\rm R}}(p)}\frac{Y^{{\rm L}}(q)}{Y^{{\rm R}}(p)},\\ & r_8^{{\rm RL}}=-i\,h_2^{{\rm R}}(p)\frac{h_2^{{\rm L}}(q)-h_1^{{\rm L}}(q)}{1+h_1^{{\rm L}}(q)h_2^{{\rm R}}(p)}\frac{X^{{\rm L}}(q)}{X^{{\rm R}}(p)}.
 \end{align}
 Finally, the left-right sector ($\tA=\tL$ and $ \tB=\tR $) is given by
 \begin{align}
 & r_1^{{\rm LR}}=1,\quad r_5^{{\rm LR}}=0=r_6^{{\rm LR}}\\
 & r_2^{{\rm LR}}=-\frac{1}{h_2^{{\rm L}}(p)-h_1^{{\rm L}}(p)}\frac{1+h_1^{{\rm L}}(p)h_1^{{\rm R}}(q)}{1+h_2^{{\rm L}}(p)h_1^{{\rm R}}(q)}\frac{1}{X^{{\rm L}}(p)Y^{{\rm L}}(p)},\\
 & r_3^{{\rm LR}}=-\frac{h_2^{{\rm R}}(q)-h_1^{{\rm R}}(q)}{h_2^{{\rm R}}(q)^2}\frac{1+h_2^{{\rm L}}(p)h_2^{{\rm R}}(q)}{1+h_2^{{\rm L}}(p)h_1^{{\rm R}}(q)}X^{{\rm R}}(q)Y^{{\rm R}}(q),\\
 & r_4^{{\rm LR}}=-\frac{1}{h_2^{{\rm R}}(q)^2}\frac{h_2^{{\rm R}}(q)-h_1^{{\rm R}}(q)}{h_2^{{\rm L}}(p)-h_1^{{\rm L}}(p)}\frac{1+h_1^{{\rm L}}(p)h_2^{{\rm R}}(q)}{1+h_2^{{\rm L}}(p)h_1^{{\rm R}}(q))}\frac{X^{{\rm R}}(p)Y^{{\rm R}}(p)}{X^{{\rm L}}(p)Y^{{\rm L}}(p)},\\
 & r_7^{{\rm LR}}=\frac{i}{h_2^{{\rm R}}(q)}\frac{h_2^{{\rm R}}(q)-h_1^{{\rm R}}(q)}{1+h_2^{{\rm L}}(p)h_1^{{\rm R}}(q)}\frac{X^{{\rm R}}(q)}{X^{{\rm L}}(p)},\\ & r_8^{{\rm LR}}=-\frac{i}{h_2^{{\rm R}}(q)}\frac{h_2^{{\rm R}}(q)-h_1^{{\rm R}}(q)}{1+h_2^{{\rm L}}(p)h_1^{{\rm R}}(q)}\frac{Y^{{\rm R}}(q)}{Y^{{\rm L}}(p)}.
 \end{align}
With these four blocks satisfying all the eight YBE$ (\tA,\tB,\tC) $, the full $ R $-matrix $ \bbR $ is given by \eqref{fullRmatrixgeneral}, and it satisfies \eqref{YBE16x16}. It is remarkable that it is possible to deform the left-left and right-right blocks independently and still obtain meaningful right-left and left-right blocks. Because of this independence of the diagonal blocks, 6-vertex B is both a deformation of AdS$_3\times\text{S}^3\times \text{T}^4$ \cite{Sfondrini:2014via,Borsato:2014exa}  and AdS$ _3\times \text{S}^3\times \text{S}^3\times \text{S}^1 $\cite{Borsato:2015mma}. More details about the comparison between these models and the undeformed ones are given in section \ref{Comparison}.

We find braiding unitarity in each of the four blocks
 \begin{align}
 & \lR^{{\rm RR}}(p,q)\lP\lR^{{\rm RR}}(q,p)\lP=\mathcal{B}^{{\rm RR}}(p,q)\lI,\\
 & \lR^{{\rm LL}}(p,q)\lP\lR^{{\rm LL}}(q,p)\lP=\mathcal{B}^{{\rm LL}}(p,q)\lI,\\
 & \lR^{{\rm RL}}(p,q)\lP\lR^{{\rm LR}}(q,p)\lP=\mathcal{B}^{{\rm RL}}(p,q)\lI,\\
& \lR^{{\rm LR}}(p,q)\lP\lR^{{\rm RL}}(q,p)\lP=\mathcal{B}^{{\rm LR}}(p,q)\lI
 \end{align}
 where 
 \begin{align}
&  \mathcal{B}^{{\rm RR}}(p,q) \hspace{-2.2cm}&&=\frac{h_2^{{\rm R}}(p)-h_1^{{\rm R}}(q)}{h_2^{{\rm R}}(p)-h_1^{{\rm R}}(p)}\frac{h_2^{{\rm R}}(q)-h_1^{{\rm R}}(p)}{h_2^{{\rm R}}(q)-h_1^{{\rm R}}(q)}\sigma^{{\rm RR}}(p,q)\sigma^{{\rm RR}}(q,p),\\
& \mathcal{B}^{{\rm LL}}(p,q)\hspace{-2.2cm}&&=\frac{h_2^{{\rm L}}(p)-h_1^{{\rm L}}(q)}{h_2^{{\rm L}}(p)-h_1^{{\rm L}}(p)}\frac{h_2^{{\rm L}}(q)-h_1^{{\rm L}}(p)}{h_2^{{\rm L}}(q)-h_1^{{\rm L}}(q)}\sigma^{{\rm LL}}(p,q)\sigma^{{\rm LL}}(q,p),\\
& \mathcal{B}^{{\rm RL}}(p,q)\hspace{-2.2cm}&&=\frac{1+h_2^{{\rm L}}(q)h_2^{{\rm R}}(p)}{1+h_1^{{\rm L}}(q)h_2^{{\rm R}}(p)}\frac{1+h_1^{{\rm R}}(q)h_1^{{\rm L}}(p)}{1+h_1^{{\rm R}}(p)h_2^{{\rm L}}(q)}\sigma^{{\rm RL}}(p,q)\sigma^{{\rm LR}}(q,p),\\
& \mathcal{B}^{{\rm LR}}(p,q)\hspace{-2.2cm}&&=\frac{1+h_2^{{\rm L}}(p)h_2^{{\rm R}}(q)}{1+h_1^{{\rm L}}(p)h_2^{{\rm R}}(q)}\frac{1+h_1^{{\rm R}}(q)h_1^{{\rm L}}(p)}{1+h_1^{{\rm R}}(q)h_2^{{\rm L}}(p)}\sigma^{{\rm LR}}(p,q)\sigma^{{\rm RL}}(q,p).
 \end{align}
 With the above expressions, if
\begin{equation}
\mathbb{B}(p,q)\equiv\mathcal{B}^{{\rm RR}}(p,q) =\mathcal{B}^{{\rm LL}}(p,q) =\mathcal{B}^{{\rm RL}}(p,q) =\mathcal{B}^{{\rm LR}}(p,q) 
\end{equation}
then the full $ R $-matrix $ \bbR $ \eqref{fullRmatrixgeneral} automatically satisfies braiding unitarity
 \begin{equation}
 \bbR(p,q)\bP\bbR(q,p)\bP=\mathbb{B}(p,q)\,\mathds{1}.
 \label{braiding6vB}
 \end{equation}

\subsection{A deformation of AdS$_3 $: 8-vertex B}\label{subsec:full8vB}

In this section we present the four $ 4\times 4 $ blocks (and consequently the full $ R $-matrix) for the 8-vertex model introduced in \cite{deLeeuw:2020ahe}. This model can be  seen as a deformation of AdS$_3\times\text{S}^3\times \text{M}^4$ $R$-matrix introduced in \cite{Sfondrini:2014via,Borsato:2014exa,Borsato:2015mma}. The following notation will be used
\begin{align}
& \eta_\pm=\frac{\eta(u)\pm\eta(v)}{2}, && \text{sn}^{\rm AB}_\pm=\text{sn}(G^{\rm A}(u)\pm G^{\rm B}(v),k^2),\\
&  \text{cn}^{\rm AB}_\pm=\text{cn}(G^{\rm A}(u)\pm G^{\rm B}(v),k^2), && \text{dn}^{\rm AB}_\pm=\text{dn}(G^{\rm A}(u)\pm G^{\rm B}(v),k^2). 
\label{notationJacobi}
\end{align}
 The explicit form of each matrix element $ r_i^{{\rm AB}} $ is presented below, starting by the left-left block
\begin{align}
&r_1^{{\rm LL}}=\frac{1}{\sqrt{\sin(\eta(u))\sin(\eta(v))}}\left(-\cos\eta_+\text{sn}^{{\rm LL}}_-+\frac{\text{cn}^{{\rm LL}}_-}{\text{dn}^{{\rm LL}}_-}\sin\eta_+\right),\nonumber\\
&r_2^{{\rm LL}}=-\frac{1}{\sqrt{\sin(\eta(u))\sin(\eta(v))}}\left(\cos\eta_-\text{sn}^{{\rm LL}}_--\frac{\text{cn}^{{\rm LL}}_-}{\text{dn}^{{\rm LL}}_-}\sin\eta_-\right),\nonumber\\
&r_3^{{\rm LL}}=-\frac{1}{\sqrt{\sin(\eta(u))\sin(\eta(v))}}\left(\cos\eta_-\text{sn}^{{\rm LL}}_--\frac{\text{cn}^{{\rm LL}}_-}{\text{dn}^{{\rm LL}}_-}\sin\eta_-\right),\nonumber\\
&r_4^{{\rm LL}}=\frac{1}{\sqrt{\sin(\eta(u))\sin(\eta(v))}}\left(\cos\eta_+\text{sn}^{{\rm LL}}_-+\frac{\text{cn}^{{\rm LL}}_-}{\text{dn}^{{\rm LL}}_-}\sin\eta_+\right),\nonumber\\
&r_5^{{\rm LL}}=\sqrt{\frac{g_{{\rm L}}(v)}{g_{{\rm L}}(u)}}, \quad r_6^{{\rm LL}}=\sqrt{\frac{g_{{\rm L}}(u)}{g_{{\rm L}}(v)}},\nonumber\\
& r_7^{{\rm LL}}=\frac{k \alpha}{\sqrt{g_{{\rm L}}(u)g_{{\rm L}}(v)}}\frac{\text{cn}^{{\rm LL}}_-\text{sn}^{{\rm LL}}_-}{\text{dn}^{{\rm LL}}_-}, \quad r_8^{{\rm LL}}=\frac{k \sqrt{g_{{\rm L}}(u)g_{{\rm L}}(v)}}{\alpha}\frac{\text{cn}^{{\rm LL}}_-\text{sn}^{{\rm LL}}_-}{\text{dn}^{{\rm LL}}_-}.
\label{eq:LLblock8vB}
\end{align}
This is simply the 8-vertex B $R$-matrix where we added a diagonal local basis transformation with component $g_{{\rm L}}$.
This block satisfies the YBE$ (\tL,\tL,\tL) $ (equation \eqref{YBE8eqs}). 

Similarly, let us now introduce the right-right block
\begin{align}
&r_1^{{\rm RR}}=\frac{1}{\sqrt{\sin(\eta(u))\sin(\eta(v))}}\left(-\cos\eta_+\text{sn}^{{\rm RR}}_-+\frac{\text{cn}^{{\rm RR}}_-}{\text{dn}^{{\rm RR}}_-}\sin\eta_+\right),\nonumber\\
&r_2^{{\rm RR}}=-\frac{1}{\sqrt{\sin(\eta(u))\sin(\eta(v))}}\left(\cos\eta_-\text{sn}^{{\rm RR}}_-+\frac{\text{cn}^{{\rm RR}}_-}{\text{dn}^{{\rm RR}}_-}\sin\eta_-\right),\nonumber\\
&r_3^{{\rm RR}}=-\frac{1}{\sqrt{\sin(\eta(u))\sin(\eta(v))}}\left(\cos\eta_-\text{sn}^{{\rm RR}}_--\frac{\text{cn}^{{\rm RR}}_-}{\text{dn}^{{\rm RR}}_-}\sin\eta_-\right),\nonumber\\
&r_4^{{\rm RR}}=\frac{1}{\sqrt{\sin(\eta(u))\sin(\eta(v))}}\left(\cos\eta_+\text{sn}^{{\rm RR}}_-+\frac{\text{cn}^{{\rm RR}}_-}{\text{dn}^{{\rm RR}}_-}\sin\eta_+\right),\nonumber\\
&r_5^{{\rm RR}}=\sqrt{\frac{g_{{\rm R}}(u)}{g_{{\rm R}}(v)}}, \quad r_6^{{\rm LL}}=\sqrt{\frac{g_{{\rm R}}(v)}{g_{{\rm R}}(u)}},\nonumber\\
& r_7^{{\rm RR}}=\frac{k \sqrt{g_{{\rm R}}(u)g_{{\rm R}}(v)}}{\alpha}\frac{\text{cn}^{{\rm RR}}_-\text{sn}^{{\rm RR}}_-}{\text{dn}^{{\rm RR}}_-}, \quad r_8^{{\rm RR}}=\frac{k \alpha}{\sqrt{g_{{\rm R}}(u)g_{{\rm R}}(v)}}\frac{\text{cn}^{{\rm RR}}_-\text{sn}^{{\rm RR}}_-}{\text{dn}^{{\rm RR}}_-},
\label{eq:RRblock8vB}
\end{align}
that satisfies the YBE$ (\tR,\tR,\tR) $. Notice that \textit{a priori} $\lR^{{\rm LL}}(u,v)$ and $\lR^{{\rm RR}}(u,v)$ can have different elliptic moduli $k_L,k_R$ and different functions $\eta_{\mathrm{L,R}}$. However, by computing the $ {\rm LR} $ and $ {\rm RL} $ blocks we find that the parameters need to be related and the same is true for the function $\eta$. Indeed, by using different $\eta_{\rm L}$ and $\eta_{\rm R}$ in each block we found that they are related by $\eta_R = \pi-\eta_L$. 

Using the method described in Section \ref{subsec:generalprocedure} we can then construct the following right-left block
\begin{align}
& r_1^{{\rm RL}}=\frac{1}{\sqrt{\sin(\eta(u))\sin(\eta(v))}}\sqrt{\frac{g_{{\rm R}}(u)}{g_{{\rm L}}(v)}}\left(\cos\eta_-\text{sn}^{{\rm RL}}_++\frac{\text{cn}^{{\rm RL}}_+}{\text{dn}^{{\rm RL}}_+}\sin\eta_-\right),\nonumber\\
& r_2^{{\rm RL}}=\frac{1}{\sqrt{\sin(\eta(u))\sin(\eta(v))}}\sqrt{\frac{g_{{\rm R}}(u)}{g_{{\rm L}}(v)}}\left(\cos\eta_+\text{sn}^{{\rm RL}}_+-\frac{\text{cn}^{{\rm RL}}_+}{\text{dn}^{{\rm RL}}_+}\sin\eta_+\right),\nonumber\\
&r_3^{{\rm RL}}=\frac{1}{\sqrt{\sin(\eta(u))\sin(\eta(v))}}\sqrt{\frac{g_{{\rm R}}(u)}{g_{{\rm L}}(v)}}\left(\cos\eta_+\text{sn}^{{\rm RL}}_++\frac{\text{cn}^{{\rm RL}}_+}{\text{dn}^{{\rm RL}}_+}\sin\eta_+\right),\nonumber\\
&r_4^{{\rm RL}}=\frac{1}{\sqrt{\sin(\eta(u))\sin(\eta(v))}}\sqrt{\frac{g_{{\rm R}}(u)}{g_{{\rm L}}(v)}}\left(-\cos\eta_-\text{sn}^{{\rm RL}}_++\frac{\text{cn}^{{\rm RL}}_+}{\text{dn}^{{\rm RL}}_+}\sin\eta_-\right),\nonumber\\
& r_5^{{\rm RL}}=-\frac{k g_{{\rm R}}(u)}{\alpha}\frac{\text{cn}^{{\rm RL}}_+\text{sn}^{{\rm RL}}_+}{\text{dn}^{{\rm RL}}_+},\quad r_6^{{\rm RL}}=\frac{k \alpha}{g_{{\rm L}}(v)}\frac{\text{cn}^{{\rm RL}}_+\text{sn}^{{\rm RL}}_+}{\text{dn}^{{\rm RL}}_+},\nonumber\\
&r_7^{{\rm RL}}=-\frac{g_{{\rm R}}(u)}{g_{{\rm L}}(v)},\quad r_8^{{\rm RL}}=1,
\label{eq:RLblock8vB}
\end{align}
and finally the left-right block
\begin{align}
& r_1^{{\rm LR}}=\frac{1}{\sqrt{\sin(\eta(u))\sin(\eta(v))}}\sqrt{\frac{g_{{\rm R}}(v)}{g_{{\rm L}}(u)}}\left(\cos\eta_-\text{sn}^{{\rm LR}}_++\frac{\text{cn}^{{\rm LR}}_+}{\text{dn}^{{\rm LR}}_+}\sin\eta_-\right),\nonumber\\
& r_2^{{\rm LR}}=\frac{1}{\sqrt{\sin(\eta(u))\sin(\eta(v))}}\sqrt{\frac{g_{{\rm R}}(v)}{g_{{\rm L}}(u)}}\left(\cos\eta_+\text{sn}^{{\rm LR}}_+-\frac{\text{cn}^{{\rm LR}}_+}{\text{dn}^{{\rm LR}}_+}\sin\eta_+\right),\nonumber\\
&r_3^{{\rm LR}}=\frac{1}{\sqrt{\sin(\eta(u))\sin(\eta(v))}}\sqrt{\frac{g_{{\rm R}}(v)}{g_{{\rm L}}(u)}}\left(\cos\eta_+\text{sn}^{{\rm LR}}_++\frac{\text{cn}^{{\rm LR}}_+}{\text{dn}^{{\rm LR}}_+}\sin\eta_+\right),\nonumber\\
&r_4^{{\rm LR}}=\frac{1}{\sqrt{\sin(\eta(u))\sin(\eta(v))}}\sqrt{\frac{g_{{\rm R}}(v)}{g_{{\rm L}}(u)}}\left(-\cos\eta_-\text{sn}^{{\rm LR}}_++\frac{\text{cn}^{{\rm LR}}_+}{\text{dn}^{{\rm LR}}_+}\sin\eta_-\right),\nonumber\\
& r_5^{{\rm LR}}=-\frac{k \alpha}{g_{{\rm L}}(u)}\frac{\text{cn}^{{\rm LR}}_+\text{sn}^{{\rm LR}}_+}{\text{dn}^{{\rm LR}}_+},\quad r_6^{{\rm LR}}=\frac{k g_{{\rm R}}(v)}{\alpha}\frac{\text{cn}^{{\rm LR}}_+\text{sn}^{{\rm LR}}_+}{\text{dn}^{{\rm LR}}_+},\nonumber\\
&r_7^{{\rm LR}}=-\frac{g_{{\rm R}}(u)}{g_{{\rm L}}(v)},\quad r_8^{{\rm LR}}=1.
\label{eq:LRblock8vB}
\end{align}
We can now immediately construct the full $ R $-matrix $ \bbR $  \eqref{fullRmatrixgeneral} and check that it indeed satisfies the Yang-Baxter equation \eqref{YBE16x16}.

Now let us discuss braiding unitarity. For the four blocks just presented we have
\begin{align}
& \lR^{{\rm RL}}(u,v)\lP\lR^{{\rm LR}}(v,u)\lP=\mathcal{B}^{{\rm RL}}(u,v)\lI,\\
& \lR^{{\rm LR}}(u,v)\lP\lR^{{\rm RL}}(v,u)\lP=\mathcal{B}^{{\rm LR}}(u,v)\lI,\\
& \lR^{{\rm RR}}(u,v)\lP\lR^{{\rm RR}}(v,u)\lP=\mathcal{B}^{{\rm RR}}(u,v)\lI,\\
& \lR^{{\rm LL}}(u,v)\lP\lR^{{\rm LL}}(v,u)\lP=\mathcal{B}^{{\rm LL}}(u,v)\lI.
\end{align}
where 
\begin{align}
\frac{\mathcal{B}^{{\rm LL}}(u,v)}{\sigma^{{\rm LL}}(u,v)\sigma^{{\rm LL}}(v,u)} &= \frac{\text{cn}_{\text{L,L},-}^2}{\text{dn}_{\text{L,L},-}^2} \left(-\frac{\sin^2 \eta_+}{\sin \eta(u) \sin \eta(v)} +k^2 \text{sn}_{\text{L,L},-}^2\right) - \text{sn}_{\text{L,L},-}^{2} \frac{\cos ^2\eta_+}{\sin \eta(u) \sin \eta(v)} \\
\frac{\mathcal{B}^{{\rm RR}}(u,v)}{ \sigma^{{\rm RR}}(u,v)\sigma^{{\rm RR}}(v,u)}& = \frac{\text{cn}_{\text{R,R},-}^2}{\text{dn}_{\text{R,R},-}^2} \left(-\frac{\sin^2 \eta_+}{\sin \eta(u) \sin \eta(v)} +k^2 \text{sn}_{\text{R,R},-}^2\right) - \text{sn}_{\text{R,R},-}^{2} \frac{\cos ^2\eta_+}{\sin \eta(u) \sin \eta(v)}
\end{align}
\begin{align}
\frac{g^{{\rm R}}(v)}{g^{{\rm L}}(u)}\frac{\mathcal{B}^{{\rm LR}}(u,v)}{\sigma^{{\rm LR}}(u,v)\sigma^{{\rm RL}}(v,u)}&=-\frac{\text{cn}_{\text{L,R},+}^2}{\text{dn}_{\text{L,R},+}^2} \frac{\sin^2 \eta_-}{\sin \eta(u) \sin \eta(v)}  + \text{sn}_{\text{L,R},+}^2 \frac{\cos ^2\eta_-}{\sin \eta(u) \sin \eta(v)}-1 \\
\frac{g^{{\rm R}}(u)}{g^{{\rm L}}(v)}\frac{\mathcal{B}^{{\rm RL}}(u,v)}{\sigma^{{\rm RL}}(u,v)\sigma^{{\rm LR}}(v,u)}&=-\frac{\text{cn}_{\text{R,L},+}^2}{\text{dn}_{\text{R,L},+}^2} \frac{\sin^2 \eta_-}{\sin \eta(u) \sin \eta(v)}  + \text{sn}_{\text{R,L},+}^2 \frac{\cos ^2\eta_-}{\sin \eta(u) \sin \eta(v)}-1, 
\end{align}
where $ \text{cn}_{A,B,\pm}=\text{cn}(G^A(u)\pm G^B(v),k^2) $, and similarly for dn and sn.

In order for the full $ R $-matrix $ \bbR(u,v) $ to satisfy braiding unitarity 
\begin{equation}
\bbR(u,v)\bP\bbR(v,u)\bP=\mathbb{B}(u,v)\bI
\end{equation}
it is necessary that
\begin{equation}
\mathbb{B}(u,v)\equiv \mathcal{B}^{{\rm LL}}(u,v)=\mathcal{B}^{{\rm RR}}(u,v)=\mathcal{B}^{{\rm RL}}(u,v)=\mathcal{B}^{{\rm LR}}(u,v).
\end{equation}
This imposes additional constraints on $\sigma^{AB}$.

\section{Embeddings of $ AdS_3 \times S^{3} \times \mathrm{M}^{4} $}\label{Comparison}

Let us now show how to precisely embed the various AdS$_3$ $R$-matrices into the general ones that we derived in the previous section.

\subsection{Recovering $\mathrm{AdS}_3\times \mathrm{S}^3\times \mathrm{S}^3\times \mathrm{S}^1$ $R$-matrix from 6-vertex B }

Now we compare the 6 vertex B full $ R $-matrix $ \bbR(u,v) $ with the $ R $-matrix for AdS$ _3 \times S^3\times S^3 \times S^1$ \cite{Borsato:2015mma} given in section \ref{AdS3background2} and we obtain 
\begin{align}
& h_1^{{\rm R}}(p)=-\frac{x_{{\rm R}}^-(p)}{\beta} && h_1^{{\rm L}}(p)=\beta\, x_{{\rm L}}^-(p),\\
& h_2^{{\rm R}}(p)=-\frac{x_{{\rm R}}^+(p)}{\beta} && h_2^{{\rm L}}(p)=\beta\, x_{{\rm L}}^+(p),
\end{align}
where $ \beta $ is an arbitrary constant.

Also
\begin{align}
& X^{{\rm L}}(p)=\frac{\rho}{\gamma^{{\rm L}}(p)},&&
 Y^{{\rm L}}(p)=\frac{\gamma^{{\rm L}}(p)}{\beta\rho\left(x_{{\rm L}}^-(p)-x_{{\rm L}}^+(p)\right)}\sqrt{\frac{x_{{\rm L}}^-(p)}{x_{{\rm L}}^+(p)}},\\
& X^{{\rm R}}(p)=-i\,\rho\,\frac{ x_{{\rm R}}^+(p)}{\gamma^{{\rm R}}(p)},&&
Y^{{\rm R}}(p)=\frac{-i\,\gamma^{{\rm L}}(p)}{\beta\,\rho}\frac{\sqrt{x_{{\rm R}}^-(p)x_{{\rm R}}^+(p)}}{x_{{\rm R}}^-(p)-x_{{\rm R}}^+(p)},
\end{align}
where $ \rho $ is an arbitrary constant; and 
\begin{align}
& \sigma^{{\rm LL}}(p,q)=\frac{x_{{\rm L}}^+(p)-x_{{\rm L}}^-(p)}{x_{{\rm L}}^+(q)-x_{{\rm L}}^-(p)}\\ 
&\sigma^{{\rm RR}}(p,q)=\frac{x_{{\rm R}}^+(p)-x_{{\rm R}}^-(p)}{x_{{\rm R}}^+(q)-x_{{\rm R}}^-(p)}\\ 
&\sigma^{{\rm LR}}(p,q)=\sqrt{\frac{x_{{\rm L}}^-(p)}{x_{{\rm L}}^+(p)}}\frac{1-x_{{\rm L}}^+(p)x_{{\rm R}}^-(q)}{1-x_{{\rm L}}^-(p)x_{{\rm R}}^-(q)}\zeta^{{\rm LR}}(p,q),\\
& \sigma^{{\rm RL}}(p,q)=\sqrt{\frac{x_{{\rm R}}^-(p)}{x_{{\rm R}}^+(p)}}\frac{1-x_{{\rm R}}^+(p)x_{{\rm L}}^-(q)}{1-x_{{\rm R}}^-(p)x_{{\rm L}}^-(q)}\zeta^{{\rm RL}}(p,q).
\end{align}
Notice that the left and right sectors have their own Zhukovsky variables. As explained in Appendix A, the $\mathrm{AdS}_3\times \mathrm{S}^3 \times\mathrm{T}^4$ is a special case of this where the $x^\pm$ variables in both sectors coincide.

\subsection{Recovering the AdS$ _3 $ q-deformation from 6-vertex B}

The $R$-matrix in \cite{Hoare:2014oua,Seibold:2021lju} can be obtained from the full 6vB $ R $-matrix by making the following identifications:

\begin{align}
& h_1^{{\rm R}}(p)=-\frac{x_R^-(p)}{\beta}, && h_1^{{\rm L}}(p)=\beta\,x_L^-(p), \label{eq:qdefIdent1}\\
& h_2^{{\rm R}}(p)=-\frac{x_R^+(p)}{\beta}, && h_2^{{\rm L}}(p)=\beta\,x_L^+(p),
\end{align}
where $ \beta  $ is an arbitrary constant.

Also, 
\begin{align}
& X^{{\rm L}}(p)=\frac{\rho}{\gamma_L(p)}, && Y^{{\rm L}}(p)=\frac{1}{\beta\,\rho}\frac{\gamma_L(p)}{U_L(p)V_L(p)W_L(p)}\frac{1}{x_L^-(p)-x_L^+(p)},\\
& Y^{{\rm R}}(p)=\frac{1}{\beta\,\rho}\frac{x_R^+(p)}{\gamma_R(p)}, && X^{{\rm R}}(p)=-\frac{\rho\,\gamma_R(p)}{U_R(p)V_R(p)W_R(p)}\frac{x_R^+(p)}{x_R^-(p)-x_R^+(p)}\label{eq:qdefIdent4},
\end{align}
where $ \rho $ is an arbitrary constant; and
\begin{align}
& \sigma^{{\rm LL}}(p,q)=-\frac{U_L(p)V_L(p)W_L(p)}{U_L(q)V_L(q)W_L(q)}\frac{x_L^-(p)-x_L^+(p)}{x_L^-(q)-x_L^+(p)},\\
&\sigma^{{\rm RR}}(p,q)=-\frac{U_R(p)V_R(p)W_R(p)}{U_R(q)V_R(q)W_R(q)}\frac{x_R^-(p)-x_R^+(p)}{x_R^-(q)-x_R^+(p)},\\
&\sigma^{{\rm LR}}(p,q)=U_R(q)V_R(q)W_R(q)\frac{(1-x_L^+(p)x_R^-(q))}{(1-x_L^+(p)x_R^+(q))},\\
&\sigma^{{\rm RL}}(p,q)=U_L(q)V_L(q)W_L(q)\frac{(1-x_R^+(p)x_L^-(q))}{(1-x_R^+(p)x_L^+(q))}.
\end{align}
In particular, notice that the identifications \eqref{eq:qdefIdent1}-\eqref{eq:qdefIdent4} are invertible. This means that this $q$-deformation is the most general deformation possible. The only source of further deformations comes from the fact that our functions are completely unconstraint. This basically translates in making the constants in the $q$-deformed model, such as the mass, dependent on the spectral parameter. This is what we call a functional deformation.

\subsection{Recovering $\mathrm{AdS}_3\times \mathrm{S}^3\times \mathrm{S}^3\times \mathrm{S}^1$ $R$-matrix from 8-vertex B}

In order to recover the $\mathrm{AdS}_3\times \mathrm{S}^3\times \mathrm{S}^3\times \mathrm{S}^1$ $ R $-matrix we need to take the limit $ k\rightarrow 0 $ in the full 8-vertex $R$-matrix as presented in section \ref{subsec:full8vB} and compare it with \cite{Sfondrini:2014via,Borsato:2014exa} which is given in Appendix \ref{AdS3background2} .

After the limit $ k\rightarrow 0 $, the map to recover $\mathrm{AdS}_3\times \mathrm{S}^3\times \mathrm{S}^3\times \mathrm{S}^1$ $ R $-matrix is the following:
\begin{align}
&\sigma^{\text{LR}}(p,q)=\sqrt{\frac{x_R^-(q)}{x_R^+(q)}}\frac{x_L^-(p)-x_L^+(p)}{1-x_L^-(p)x_R^-(q)}\frac{\gamma^R(q)}{\gamma^L(p)},\\
&\sigma^{\text{RL}}(p,q)=\sqrt{\frac{x_R^-(p)}{x_R^+(p)}}\frac{x_L^-(p)-x_L^+(p)}{1-x_L^-(q)x_R^-(p)}\frac{\gamma^R(p)}{\gamma^L(q)},\\
&\sigma^{\text{LL}}(p,q)=\frac{\sqrt{g_\text{L}(q)}}{\sqrt{g_\text{L}(p)}}\frac{x_L^-(p)-x_L^+(p)}{x_L^-(p)-x_L^+(q)}\frac{\gamma^L(q)}{\gamma^L(p)},\\
&\sigma^{\text{RR}}(p,q)=\frac{\sqrt{g_\text{R}(q)}}{\sqrt{g_\text{R}(p)}}\frac{x_R^-(p)-x_R^+(p)}{x_R^-(p)-x_R^+(q)}\frac{\gamma^R(q)}{\gamma^R(p)}.
\end{align}
 Moreover,
\begin{equation}
g_{\text{L}}(p)=\tau \frac{x_L^-(p)-x_L^+(p)}{\gamma^L(p)^2}\sqrt{\frac{x_L^+(p)}{x_L^-(p)}} \quad \text{and}\quad  g_{\text{R}}(p)=\tau\frac{x_R^-(p)-x_R^+(p)}{\gamma^R(p)^2}\sqrt{\frac{x_R^+(p)}{x_R^-(p)}} ,
\end{equation}
and
\begin{align}
& G^L(p)=\pi -\frac{i}{4}\log\left(\sqrt{x_L^-(p)x_L^+(p)}\right) \\
& G^R(p)=\pi -\frac{i}{4}\log\left(\sqrt{x_R^-(p)x_R^+(p)}\right) \\
& \eta(p)= -\frac{i}{2} \log\left(\sqrt{\frac{x_L^+(p)}{x_L^-(p)}}\right),
\end{align}
where $ \tau=\pm 1 $. This identification only works since $\frac{x_L^+}{x_L^-}= \frac{x_R^+}{x_R^-}$. 

 
\section{Deformed symmetry algebras}\label{symmetryalgs}

Let us now try to get some insight into the physical interpretation of the deformations that we have derived. The natural starting point for this is to consider the symmetry algebras that these models exhibit. We will determine the algebra generated by generators $x$ such that quasi-cocommutativity is satisfied
\begin{equation}\label{opcoop}
\Delta^{\rm op}(x)R = R \Delta(x)\,.
\end{equation}

\subsection{General procedure}\label{symproc}

For a given $R$-matrix, the symmetry algebra is generated by the so-called RTT-relations
\begin{align}
R_{12}(u,v) T_{1}(u) T_{2}(v) =  T_{2}(v)T_{1}(u) R_{12}(u,v).
\label{RTT}
\end{align}
The $T$-matrices are matrices whose entries are the formal generators of the algebra and the above relations describe the fundamental commutation relations between them. Usually these relations describe algebras such as Yangian or quantum affine algebras corresponding to some underlying Lie algebra \cite{Drinfeld:1985rx,Drinfeld1986quantum}. The procedure for dealing with the RTT-realization was first formulated in \cite{Beisert:2014hya} for the $\mathrm{AdS}_5\times\mathrm{S}^5$ $S$-matrix and the Hubbard model and was later applied to $ \mathrm{AdS}_{2,3} $ \cite{Pittelli:2014ria,Hoare:2015kla,Pittelli:2017spf}.

The symmetry relations can be determined by expanding $T$ around some point where $R$ becomes (almost) the identity operator. For clarity, let us assume that this point is at $u=\infty$.
We write 
\begin{equation}
R(u,v)=R^{i_1 i_2}_{j_1 j_2} E_{i_1 j_1}\otimes E_{i_2 j_2}
\end{equation}
and 
\begin{equation}
T_a(u)=E_{ij}\otimes 1\otimes T^i_{\ j}(u),\quad T_b(u)=E_{ij}\otimes T^i_{\ j}(u)\otimes 1,
\end{equation}
then write \eqref{RTT}
in component form. Next, we write an expansion of $T$ as 
\begin{equation}
T^{a}_{\ b}(u) =T_{(-1)b}^{\ \ a} + \sum_{s=0}^{\infty} T_{(s)b}^{\ a}\,u^{-1-s}.
\end{equation}
Then, it is easy to see that the RTT-relations reduce to regular commutation relations of some Lie algebra generated by the generators $T^a{}_{(0)b}$. The component $T^a{}_{(-1)b} \equiv \delta_{ab} \kU_a$ gives rise to the braiding element \cite{Beisert:2014hya} of the underlying coproduct which is also completely fixed.

Now, remarkably, the $R$-matrix does not only inherently contain the symmetry algebra, but also (part of) the representation theory of this algebra. Indeed, by considering the map
\begin{align}
\rho^F : T_a(u) \mapsto R_{ab}(\theta,u)
\end{align}
and using the Yang-Baxter equation we see that the components of the $R$-matrix provides the defining representation of the symmetry algebra. By a fusion procedure, more general representations can then also be constructed \cite{Beisert:2015msa}. 

Conversely, the $R$-matrix is the unique object that intertwines the coproduct and opposite coproduct in the defining representation of the symmetry algebra (including higher-order generators). This is actually the method that is used in holographic integrable models. The symmetry algebra is determined by direct computation and then the scattering matrix is derived from this. In our situation, however, we already have the $R$-matrix and can now construct the corresponding symmetry algebra and compare this again with the symmetry algebras and representations of the integrable models coming from AdS${}_{2,3}$ superstring sigma models.

\subsection{Alternative direct calculation}

The standard procedure of expanding the $R$-matrix around a point where it becomes diagonal crucially relies on the existence of such a point. In fact, for the specific case of the 8-vertex deformation of AdS$_3$  it is not possible to find such a point, despite the fact that the $R$-matrix is related to that of AdS$_2$, for which it does work.

In light of this, we will extract the (defining representation of the) symmetry algebra directly from the 8vB $R$-matrix in a manner which does not depend on the specific model at hand (i.e. on the choice of free functions) nor on the existence of a point where the $R$-matrix becomes diagonal. We can then fine tune the obtained algebra to the model at hand. 

Consider some symmetry generator $\kQ$. We put $\check{R}(u,v) = \mathcal{P} R(u,v)$ where $\mathcal{P}$ is the permutation operator and define 
\begin{equation}
\kQ_{12}(u,v)=\kQ(u) \otimes 1 + \kU(u)  \otimes \kQ(v),
\end{equation}
where the tensor product is graded when we have both bosonic and fermionic degrees of freedom in our representation. Then the symmetry algebra relation \eqref{opcoop} is that
\begin{equation}
\kQ_{12}(v,u) \check{R}(u,v)  =\check{R}(u,v) \kQ_{12}(u,v).
\end{equation}
We can then differentiate this equation with respect to $u$ and subsequently put $u\rightarrow v$, which leaves us with a set of ODEs 
\begin{align}\label{eq:symH}
[\kQ \otimes 1 + \kU  \otimes \kQ ,\lH]  =  \kQ' \otimes 1 + \kU'  \otimes \kQ - \kU \otimes \kQ'\,.
\end{align}
which can be solved directly and we have introduced the Hamiltonian density $\lH(v) = \partial_u \left.\check{R}(u,v)\right|_{u\rightarrow v}$. Notice that this equation greatly resembles the Sutherland equation that is crucial in the boost operator formalism of \cite{deLeeuw:2020ahe}. Moreover, we  also see here that the symmetry algebra is fixed by the Hamiltonian density $\lH$ of the system which nicely ties in to the bottom-up approach of our classification method \cite{deLeeuw:2020ahe}.

Our approach will be to simply solve \eqref{eq:symH} for the derivatives of the remaining functions and plug them back in, obtaining a set of algebraic equations for the functions themselves. This approach is especially useful if the the $R$-matrix does not have a nice asymptotic behaviour. 

\subsection{6-vertex deformation of AdS$_3$}

Let us now consider the 6-vertex deformation of the full AdS$_3$ $R$-matrices. We will start by considering two supercharges $\kQ_\pm$ which we assume to have the form
\begin{equation}
\kQ_\pm = \left( 
\begin{array}{cc}
\kQ_\pm^{\rm L} & 0 \\
0 & \kQ_\pm^{\rm R} 
\end{array}
\right)
\end{equation}
with
\begin{equation}
\kQ_+^{\rm L} =  \left( 
\begin{array}{cc}
0 & 0 \\
a_+ & 0
\end{array}
\right),\quad \kQ_-^{\rm L} =  \left( 
\begin{array}{cc}
0 & a_- \\
0 & 0
\end{array}
\right)
\end{equation}
\begin{equation}
\kQ_+^{\rm R} =  \left( 
\begin{array}{cc}
0 & b_+ \\
0 & 0
\end{array}
\right),\quad \kQ_-^{\rm R} =  \left( 
\begin{array}{cc}
0 & 0 \\
b_- & 0
\end{array}
\right)\,.
\end{equation}
We will equip these charges with coproducts $\Delta(\kQ_\pm)$ which we assume to have the form
together with coproducts of the following form
\begin{equation}
\Delta(\kQ_+)=\kQ_+\otimes 1 + \kU^{-1} \otimes\kQ_+,\quad \Delta(\kQ_-)=\kQ_-\otimes 1 + \mathfrak{V}^{-1} \otimes\kQ_-
\end{equation}
where $\kU$ and $\mathfrak{V}$ are block diagonal matrices acting as 
\begin{equation}
\kU = \left( 
\begin{array}{cc}
\kU_{\rm L} & 0 \\
0 & \kU_{\rm R}
\end{array}
\right),\quad 
\kV = \left( 
\begin{array}{cc}
\kV_{\rm L} & 0 \\
0 & \kV_{\rm R}
\end{array}
\right)
\end{equation}
where each of $\kU_{\rm L,R}$ and $\kV_{\rm L,R}$ are scalar multiples of the identity operator.

By imposing the required commutation relations between the supercharges and the $R$-matrix we easily obtain 
\begin{equation}
\begin{split}
& a_+ =\frac{1}{X^{\rm L}}\frac{1}{t-h_2^{\rm L}},\quad a_- = X^{\rm L} \frac{h_1^{\rm L}-h_2^{\rm L}}{s-h_1^{\rm L}}\\
& b_+ = i\, \frac{Y^{\rm R}}{h_2^{\rm R}}\frac{h_1^{\rm R}-h_2^{\rm R}}{1+t\, h_1^{\rm R}},\quad b_- = i\frac{h_2^{\rm R}}{Y^{\rm R}}\frac{1}{1+s\, h_2^{\rm R}}
\end{split}
\end{equation}
where $s$ and $t$ are arbitrary constants and 
\begin{equation}
\begin{split}
& \kU_{\rm L} = X^{\rm L} Y^{\rm L} (h_1^{\rm L}-h_2^{\rm L})\frac{t-h_2^{\rm L}}{t-h_1^{\rm L}},\quad \kU_{\rm R}=\frac{h_2^{\rm R}}{X^{\rm R}}\frac{h_2^{\rm R}}{Y^{\rm R}}\frac{1}{h_1^{\rm R}-h_2^{\rm R}}\frac{1+t\,h_1^{\rm R}}{1+t\,h_2^{\rm R}}\\
& \kV_{\rm L}=\frac{1}{X^{\rm L}}\frac{1}{Y^{\rm L}}\frac{1}{h_1^{\rm L}-h_2^{\rm L}}\frac{s-h_1^{\rm L}}{s-h_2^{\rm L}},\quad \kV_{\rm R}=\frac{X^{\rm R}}{h_2^{\rm R}}\frac{Y^{\rm R}}{h_2^{\rm R}}(h_1^{\rm R}-h_2^{\rm R})\frac{1+s\, h_2^{\rm R}}{1+s\,h_1^{\rm R}}\,.
\end{split}
\end{equation} 
At first glance it seems we have two one-parameter families of conserved charges $\kQ_+(t)$ and $\kQ_-(s)$. However only two such charges from each family are actually independent since for example for any $T$ we can write 
\begin{equation}
\kQ_+(T) = \alpha\, \kQ_+(t_1)+\beta\, \kQ_-(t_2),\quad t_1\neq t_2
\end{equation}
and as a result there are four independent supercharges $\kQ_+(t_1)$, $\kQ_+(t_2)$, $\kQ_-(s_1)$, $\kQ_-(s_2)$.
The logic can then be reversed -- starting from these supercharges the $R$-matrix is completely constrained up to normalisation in each of the four blocks by the requirement 
\begin{equation}
\Delta^{\rm op}(\kQ)R = R \Delta(\kQ)\,.
\end{equation}
We will now examine the resulting algebra more closely. We have the following commutation relations
\begin{equation}
\{\kQ_+(t_1),\kQ_+(t_2)\}=0,\quad \{\kQ_-(s_1),\kQ_-(s_2)\}=0\,.
\end{equation}
We introduce a two-parameter family of operators $\kP(t,s)$ defined by 
\begin{equation}
\kP(t,s)=\{\kQ_+(t),\kQ_-(s) \}\,.
\end{equation}
Clearly there are four independent such charges and all of them are central elements of the algebra. Their coproducts are given by 
\begin{equation}
\Delta(\kP) = \kP\otimes 1 + \kV^{-1}\kU^{-1} \otimes \kP\,.
\end{equation}
The coproducts of the braiding factors are given by $\Delta(\kU)=\kU\otimes \kU$ and $\Delta(\kV)=\kV\otimes \kV$. 

Finally, the algebra can be extended by an additional $\mathbb{C}$ factor $\mathfrak{B}$ with the commutation relations 
\begin{equation}
[\mathfrak{B},\kQ_\pm]=\pm 2 \kQ_\pm
\end{equation}
and with trivial coproduct.

\subsection{Symmetries of deformed AdS$_2$}

Let us now work out the symmetries of the different deformed models.

\subsubsection*{Direct approach and defining representation}

As with the ${\rm AdS}_3$ model above we will start our considerations by considering a supercharge $\kQ$ of the form 
\begin{equation}\label{supcharge}
\kQ = \left(
\begin{array}{cc}
0 & b \\
a & 0 
\end{array}
 \right)
\end{equation}
with coproduct $\Delta(\kQ)=\kQ\otimes 1 + \kU \otimes \kQ$. Imposing as usual that $\Delta^{\rm op}(\kQ)R=R \Delta(\kQ)$ leads to the following constraints 
\begin{equation}
\begin{split}
& a = \frac{b}{\omega_1\omega_2}\frac{\omega_1\,{\rm dn}(2v)-1}{{\rm sn}(2v)} \\
& \kU = \frac{\omega_1\omega_2 \cos\eta\, {\rm sn}(2v)+\omega_2 \sin\eta\, {\rm cn}(2v)}{\sin\eta + \omega_1\omega_2 {\rm sn}(2v)}
\end{split}
\end{equation}
where $\omega_1,\omega_2 \in \{-1,1\}$. From the condition $\Delta^{\rm op}(\kQ)R=R \Delta(\kQ)$ it is also possible to extract an ODE for $b$. However, it is very difficult to express the solution in a useful form. To this end, we will follow a different approach. Namely $\kP = \frac{1}{2}\{\kQ,\kQ \}$ is central and as a result we have that $\Delta^{\rm op}(\kP)=\Delta(\kP)$ which implies $\kP = \rho(1-\kU^{2})$ where $\rho$ is some irrelevant constant. This relation provides an algebraic equation linking $a,b$ and $\kU$ allowing the symmetry generators to be completely determined. Note that the mentioned equation has two solutions. However, the two solutions differ only by a sign which in turn, along with $\rho$, only affects the normalisation of $\kQ$. Hence we can set $\rho=1$ and choose whichever sign we like.

\medskip

We now comment on the constructed supercharges corresponding to the four possible pairs $(\omega_1,\omega_2)$. We first note the following inversion property 
\begin{equation}
\left.\kU\right|_{\omega_1\rightarrow -\omega_1} =\kU^{-1}
\end{equation}
for fixed $\omega_2$ easily verified by direct calculation. Next, we notice that although we seem to have four supercharges only two are actually independent since there are only two independent matrices of the form \eqref{supcharge}. For definiteness we will choose the two independent supercharges to be given by $(\omega_1,\omega_2)=(\pm 1 ,-1)$ and denote the charges as 
\begin{equation}
\kQ_\pm = \left. \kQ \right|_{\omega_1=\pm 1},\quad \omega_2=-1
\end{equation}
and hence the coproducts are given by 
\begin{equation}
\Delta(\kQ_\pm)=\kQ_\pm\otimes 1 + \kU^{\pm 1} \otimes \kQ\,.
\end{equation}

The explicit form of the supercharges is in general quite complicated involving square roots of Jacobi elliptic functions. Huge simplifications occur at $k=1$ where the elliptic functions degenerate into hyperbolic functions. The explicit solution is then given by 

\begin{equation}
\kU = \frac{{\rm sech}(2v)\sin\,\eta +{\rm tanh}(2v) \cos\,\eta}{{\rm tanh}(2v)-\sin\,\eta}
\end{equation}
\begin{equation}
a = \frac{{\rm coth}(v)^{\frac{1-\omega_1}{2}}(\cos(\frac{\eta}{2})-\omega_1 \sin(\frac{\eta}{2})\tanh(v))}{{\rm coth}(2v)\sin\eta -\omega_1}
\end{equation}
\begin{equation}
b = \frac{{\rm tanh}(v)^{\frac{1-\omega_1}{2}}{\rm cosh}(v)}{{\rm cosh}(v)\sin(\frac{\eta}{2}) -\omega_1 {\rm sinh}(v)\cos(\frac{\eta}{2})}\,.
\end{equation}

\subsubsection*{General approach and higher symmetries}

We see that the underlying Lie algebra is the same for the deformed AdS$_2$ model as for the undeformed model. Let us now look at the higher symmetry generators by working out the RTT relations similar to \cite{Hoare:2015kla}.

\paragraph{RTT} Let us assume that there is a point at which the 8vB $R$-matrix becomes diagonal. Without loss of generality, let us set this point to be at $u=0$. We expand
\begin{align}
\eta(u) = a_0 + a_1 u + a_2 u^2 + \ldots
\end{align}
It is easy to see that $R$ only becomes diagonal if $a_0 = a_1=0$, so we assume this from now on.

Next, we expand our monodromy matrix $T$ in terms of symmetry generators as
\begin{align}
T(u) =
 \begin{pmatrix}
1 & 0 \\
0 & \kU 
\end{pmatrix} + 
 \begin{pmatrix}
\frac{H+B}{2} & \kU \kQ_- \\
\kQ_+  & \kU \frac{B-H}{2}
\end{pmatrix} u + \ldots
\end{align}
and we work out the fundamental commutation relations \eqref{RTT} expanded around $u=0$.

To first order we find that $\kU$ is central, \textit{i.e.}
\begin{align}
[\kU, T(v) ] =0.
\end{align}
At second order we indeed recover the centrally extended $\mathfrak{psu}(1|1)_{\rm ce}$ algebra together with the extra central element $B$. 

Remarkably, we find that the $k$ dependence drops out of the RTT relations in the first two orders, so the commutation relations of the first level Yangian generators are also unaffected and we therefore reproduce the algebra from \cite{Hoare:2015kla}, including the secret symmetry.

\paragraph{Deformed Yangian}

Since the first two levels of our algebra are exactly the same, the question arises if the deformation is purely realized in terms of the representation. However, starting from the second level, the deformation parameter appears in the structure constants. It is not possible to absorb this $k$ dependence into redefinitions of generators of the full algebra. This seems to indicate that the algebra of this deformed model is not a standard Yangian. It would be interesting to find out exactly what the structure for the higher generators is.

\paragraph{Deformed Representation}

The fact that the Lie algebra of symmetries remains unchanged and only the representation is modified might suggest that the deformation of the AdS${}_{2}$ model could be trivially absorbed into a redefinition of the physical parameters 
\begin{align}
&{\rm e} \mapsto {\rm e}_k,&& {\rm m} \mapsto {\rm m}_k,&& {\rm p} \mapsto {\rm p}_k\,.
\end{align}
There are two ways to see that this is not the case and the deformation is non-trivial. 

The first comes from the fact that the AdS${}_{2}$ $ R $-matrix satisfies a certain linear constraint on the diagonal elements 
\begin{equation}
r_1 - r_2 - r_3 - r_4 = 0 
\end{equation} 
It can be easily checked that this condition is not satisfied for the general 8vB model except at $k\rightarrow \infty$. 

The second comes from the evaluation representation structure of the model. It can be shown that the fundamental representation of the Yangian does not admit an evaluation representation. This is in particular clear by looking at the central extensions. Since the $k$ dependence only manifests itself at second level, we see that both the central extension $\kP$ and its first level Yangian counterpart $\hat{\kP}$
\begin{align}
&2\kP = \{\kQ,\kQ\} &&\mathrm{and}
&& 2\hat{\kP} = \{\hat{\kQ},\kQ\}
\end{align}
do not depend on $k$. However we find that $\hat{\hat{\kP}} = \frac{1}{2} \{\hat{\kQ},\hat{\kQ}\}$ \textit{does} depend on $k$. This prohibits the existence of an evaluation representation unless we take the $k\rightarrow\infty$ limit. And indeed the AdS${}_{2}$ model allows evaluation representations \cite{Hoare:2015kla}. 

\subsection{8-vertex deformation of AdS$_3$}

The LL and RR blocks of the full $R$-matrix correspond to the $4\times 4$ $R$-matrices which appeared above in the ${\rm AdS}_2$ model. As such we have already classified the symmetry generators for these blocks. From the above discussion we know now that both the LL and RR scattering matrices exhibit a centrally extended $\mathfrak{psu}(1|1)$ symmetry, whose representations depend on the elliptic parameter $k$. The question is now if this extends to a symmetry algebra of the full $16\times 16$ $S$-matrix. 

Let us write a supercharge $\kQ$ as 
\begin{equation}
\kQ = \left(
\begin{array}{cc}
\kQ^{\rm L} & 0 \\
0 & \kQ^{\rm R}
\end{array}
\right)\,.
\end{equation}
The matrices $\kQ^{\rm L}$ are precisely of the form given for the ${\rm AdS}_2$ deformation -- we only need to make the replacement $v\rightarrow G^{\rm L}(v)$ in the argument of the elliptic functions. The matrices $\kQ^{\rm R}$ can then be found almost for free -- a quick check of the equations coming from $\Delta^{\rm op}(\kQ)R = R\Delta(\kQ)$ yields that the equations involving different blocks do not mix and further the equations involving the entries of the RR block are identical to those for the LL block provided we simply swap replace $G^{\rm L} \rightarrow G^{\rm R}$, $\kU^{\rm L}\rightarrow \kU^{\rm R}$, $a^{\rm L} \rightarrow b^{\rm R}$ and $b^{\rm L}\rightarrow a^{\rm R}$. The only other modification is that for $\kQ^{\rm R}$ we must also include the normalisation factor $\rho$ since we are only free to set the normalisation of one of the blocks.

A straightforward calculation then yields that there are only \textit{two} supercharges which survive for the full $R$-matrix, namely 
\begin{equation}
\left(
\begin{array}{cc}
\kQ^{\rm L}_+ & 0 \\
0 & \kQ^{\rm R}_-
\end{array}
\right),\quad \left(
\begin{array}{cc}
\kQ^{\rm L}_- & 0 \\
0 & \kQ^{\rm R}_+
\end{array}
\right)\,.
\end{equation}
As a result the symmetry algebra is given by a single copy of ${\rm psu}(1|1)_{\rm ce}$. 

The same statements made about the ${\rm AdS}_2$ model also carry over to the current setting, namely for generic $k$ the model does not admit evaluation representations and the higher order Yangian generators become deformed.

\section{Crossing symmetry}\label{crossing}

In this section we study the presence of crossing symmetry in the deformed models. We find that the usual crossing type relations hold for both the AdS$ _3 $ and AdS$ _2 $ deformations, although the explicit form of the charge conjugation matrix $ \mathcal{C} $ for the AdS$ _2 $ one is very unusual.

\subsection{6-vertex deformation of AdS$_3$}

The $ R $-matrix described in Section \ref{subsec:full6vB}  satisfies the crossing equations
\begin{align}
& \mathbb{C}_1\bbR(p+\omega,q)^{t_1}\mathbb{C}_1^{-1}=\bbR(p,q)^{-1},\nonumber\\
& \mathbb{C}_2\bbR(p,q-\omega)^{t_2}\mathbb{C}_2^{-1}=\bbR(p,q)^{-1},
\label{crossingeq}
\end{align}
where the superscript $ t_1 $ and $ t_2 $ denote transposition in the first and second vector space, respectively; $ \omega $ is the crossing parameter and the conjugation matrix $ \mathbb{C} $  is given by
\begin{equation}
\mathbb{C}=\begin{pmatrix}
0 & 0 & 1 & 0\\
0 & 0 & 0 & i\\
1 & 0 & 0 & 0\\
0 & i & 0 & 0
\end{pmatrix},
\label{crossingmatrix6vB}
\end{equation}
if and only if the following conditions are satisfied:

\

\noindent\textbf{Condition 1:}
It is necessary that $ h_i^{{\rm L/R}}(p\pm\omega) $ satisfy
\begin{align}
& h_i^{{\rm R}}(p\pm\omega)=-\frac{1}{h_i^{{\rm L}}(p)},
& h_i^{{\rm L}}(p\pm\omega)=-\frac{1}{h_i^{{\rm R}}(p)},
\label{eq:condition1a}
\end{align}
for $i=1,2$ which implies that
\begin{align}
& h_i^{{\rm A}}(p\pm 2\omega)=h_i^{{\rm A}}(p).
\label{eq:condition1b}
\end{align}

\

\noindent\textbf{Condition 2:}
The functions $X,Y$ need to satisfy
\begin{equation}
X^{{\rm A}}(p\pm 2\omega)=-X^{{\rm A}}(p) \quad \text{and} \quad Y^{{\rm A}}(p\pm 2\omega)=-Y^{{\rm A}}(p),
\label{eq:condition2a}
\end{equation}
and
\begin{equation}
X^{{\rm R}}(p)=X^{{\rm L}}(p+\omega)\quad \text{and} \quad Y^{{\rm R}}(p)=Y^{{\rm L}}(p+\omega).
\label{eq:condition2b}
\end{equation}

\

\noindent\textbf{Condition 3:}
Finally we find conditions on the scalar factors $\sigma$
\begin{align}
& \sigma^{{\rm AB}}(p+\omega,q)\sigma^{{\rm BB}}(p,q)=\frac{h_2^{{\rm B}}(p)-h_1^{{\rm B}}(p)}{h_2^{{\rm B}}(q)-h_1^{{\rm B}}(p)},\\
& \sigma^{{\rm AA}}(p+\omega,q)\sigma^{{\rm BA}}(p,q)=\frac{h_2^{{\rm B}}(p)-h_1^{{\rm B}}(p)}{h_2^{{\rm B}}(p)}\frac{1+h_1^{{\rm A}}(q)h_2^{{\rm B}}(p)}{\left(1+h_1^{{\rm B}}(p)h_1^{{\rm A}}(q)\right)\left(1+h_2^{{\rm B}}(p)h_2^{{\rm A}}(q)\right)},\\
& \sigma^{{\rm AA}}(p,q-\omega)=-h_2^{{\rm A}}(p)h_2^{{\rm B}}(q)\sigma^{{\rm BB}}(p+\omega,q),\\
& \sigma^{{\rm BA}}(p,q-\omega)=\sigma^{{\rm AB}}(p+\omega,q)
\label{eq:condition3}
\end{align}
where $ \text{A},\,\text{B}=\left\{\text{L},\text{R}\right\} $ with $ \text{A}\neq \text{B} $. In the last two equations we combined both crossing equations.

Notice that the charge conjugation matrix $ \mathbb{C} $ \eqref{crossingmatrix6vB} matches with the one in \cite{Borsato:2015mma}, while corresponds the complex conjugate of the one appearing in  \cite{Sfondrini:2014via}. If one would like the exactly matching between $ \mathbb{C}  $ in \eqref{crossingmatrix6vB} and  \cite{Sfondrini:2014via} it is enough to perform a local basis transformation of the form
\begin{align}
&\mathbb{R}\mapsto \left(\mathbb{W}\otimes \mathbb{W}\right)\, \mathbb{R}\,\left(\mathbb{W}\otimes\mathbb{W}\right)^{-1},
&& \mathbb{W} = \mathrm{diag}(1,1,1,-1),
\end{align}
on our 6-vertex B full $ R $-matrix $ \mathbb{R} $.

\subsection{8-vertex deformation of AdS$_3$}\label{sec:8vBAdS3crossing}

The full $R$-matrix for 8-vertex B, deformation of AdS$ _3 $, described in Section \ref{subsec:full8vB} satisfies crossing symmetry \eqref{crossingeq} for 
\begin{equation}
\mathbb{C}=\begin{pmatrix}
0 & 0 & 1 & 0\\
0 & 0 & 0 & i\\
1 & 0 & 0 & 0\\
0 & i & 0 & 0
\end{pmatrix},
\label{eq:crossingmatrix8vB}
\end{equation}
 given that its functions satisfy
\begin{align}
G^L(p+\omega) &=2K - G^R(p) = G^L(p-\omega),\\
G^R(p+\omega) &=2K - G^L(p) = G^R(p-\omega),\\
\eta(p+\omega)&=2\pi-\eta(p)=\eta(p-\omega),\\
\sqrt{g_{{\rm R/L}}(p+\omega)}&=\sqrt{g_{{\rm L/R}}(p)}=-\sqrt{g_{{\rm R/L}}(p-\omega)}
\end{align}
where $ K $ is the Elliptic Integral of the first kind. \footnote{One needs to be particularly careful with branch cuts for this calculation. This is the reason why we are giving the form of $ \sqrt{g_{{\rm R/L}}} $ instead of $ g_{{\rm R/L}} $ itself.}.

Moreover, the overall scalar factors $\sigma^\mathrm{AB}$ of the different blocks need to satisfy the following crossing equations \footnote{To avoid cluttered expressions we denote Jacobi elliptic functions as $\text{cn}_{\text{R,L},+}$ etc in contrast to $\text{cn}^{\text{R,L}}_+$ which was used earlier in the text.}
\begin{align}
& \sigma^{{\rm LL}}(p+\omega,q)\sigma^{{\rm RL}}(p,q)=\frac{1}{i\sqrt{\frac{g_R(p)}{g_L(q)}}\left(\frac{\text{cn}_{\text{R,L},+}^2}{\text{dn}_{\text{R,L},+}^2} \frac{\sin^2 \eta_-}{\sin \eta(p) \sin \eta(q)}  - \text{sn}_{\text{R,L},+}^2 \frac{\cos ^2\eta_-}{\sin \eta(p) \sin \eta(q)}+1\right)},\\
&\sigma^{{\rm RR}}(p+\omega,q)\sigma^{{\rm LR}}(p,q)=\frac{1}{i\sqrt{\frac{g_R(q)}{g_L(p)}}\left(\frac{\text{cn}_{\text{L,R},+}^2}{\text{dn}_{\text{L,R},+}^2} \frac{\sin^2 \eta_-}{\sin \eta(p) \sin \eta(q)}  - \text{sn}_{\text{L,R},+}^2 \frac{\cos ^2\eta_-}{\sin \eta(p) \sin \eta(q)}+1\right)},\\
& \sigma^{{\rm LR}}(p+\omega,q)\sigma^{{\rm RR}}(p,q) = \frac{1}{i\sqrt{\frac{g_R(q)}{g_R(p)}}\left(\frac{\text{cn}_{\text{R,R},-}^2}{\text{dn}_{\text{R,R},-}^2} \frac{\sin^2 \eta_-}{\sin \eta(p) \sin \eta(q)}- \text{sn}_{\text{R,R},-}^2 \frac{\cos ^2\eta_-}{\sin \eta(p) \sin \eta(q)}+1\right)}\\
& \sigma^{{\rm RL}}(p+\omega,q)\sigma^{{\rm LL}}(p,q) = \frac{1}{i\sqrt{\frac{g_L(p)}{g_L(q)}}\left(\frac{\text{cn}_{\text{L,L},-}^2}{\text{dn}_{\text{L,L},-}^2} \frac{\sin^2 \eta_-}{\sin \eta(p) \sin \eta(q)}- \text{sn}_{\text{L,L},-}^2 \frac{\cos ^2\eta_-}{\sin \eta(p) \sin \eta(q)}+1\right)}
\end{align}
and
\begin{align}
& \sigma^{{\rm LL}}(p,q-\omega)=-\sigma^{{\rm RR}}(p+\omega,q)\\
& \sigma^{{\rm RR}}(p,q-\omega)=-\sigma^{{\rm LL}}(p+\omega,q)\\
& \sigma^{{\rm RL}}(p,q-\omega)=\sigma^{{\rm LR}}(p+\omega,q)\frac{g_R(q)}{g_R(p)}\\
& \sigma^{{\rm LR}}(p,q-\omega)=\sigma^{{\rm RL}}(p+\omega,q)\frac{g_L(p)}{g_L(q)}.
\end{align}
As expected the crossing matrix \eqref{eq:crossingmatrix8vB} matches with the one in \cite{Borsato:2015mma}.

\subsection{Deformed AdS$_2$}

In this section we address the question of whether the AdS$ _2 $ deformation given by the 8-vertex B $R$-matrix \eqref{8vertex} satisfies crossing symmetry for any value of the deformation parameter $ k $. The answer is particularly interesting. If we consider the boson-fermion R-matrix we do have crossing symmetry for any value of $ k $ as it will be described below.

In order to satisfy crossing symmetry the $ R $-matrix has to satisfy
\begin{align}
& \mathcal{C}_1\mathcal{R}_{12}^{st_1}(u+\omega,v)\mathcal{C}_1^{-1}=\mathcal{R}_{12}(u,v)^{-1},\nonumber\\
& \mathcal{C}_2\mathcal{R}_{12}^{st_2}(u,v-\omega)\mathcal{C}_2^{-1}=\mathcal{R}_{12}(u,v)^{-1}
\label{crossing8vb}
\end{align}
where $ \mathcal{C} $ is the charge conjugation matrix, $ \omega $ is the crossing parameter and $ st_i $ indicates super transposition in the space $ i $. The R-matrix in equations \eqref{crossing8vb} is the boson-fermion version \footnote{Just apply the following transformation: $ r_4(u,v)\rightarrow - r_4(u,v) $, $ r_7(u,v)\rightarrow i\,r_7(u,v) $ and $ r_8(u,v)\rightarrow i\,r_8(u,v) $, keeping the remaining $ r_i(u,v) $ invariant.} of \eqref{8vertex}, but with a dressing phase $ \sigma(u,v) $  and also $ \{u,v\}\rightarrow \{G(u),G(v)\} $. 

For $ \mathcal{C} $ given by\footnote{A similar solution exists for $ i\rightarrow -i $ in $\mathcal{C}$ }

\begin{equation}
\mathcal{C}=\begin{pmatrix}
0 & 1\\
-i & 0
\end{pmatrix}
\label{CforAdS2def}
\end{equation}
equations \eqref{crossing8vb} are satisfied for

\begin{equation}
\sigma(u+\omega)\sigma(u,v)=\frac{-i}{\frac{\text{cn}^2}{\text{dn}^2}\frac{\sin^2\eta_-}{\sin\eta(u)\sin\eta(v)}-\text{sn}^2\frac{\cos^2\eta_-}{\sin\eta(u)\sin\eta(v)}+1}=\sigma(u-\omega)\sigma(u,v)\label{sigmacrossAdS2}
\end{equation}
and
\begin{align}
&\eta(u+\omega)=-\eta(u)+2\pi n, &&\quad \eta(u-\omega)=-\eta(u)+2\pi m,\nonumber\\
&G(u+\omega)=G(u)+2\, n\,K(k^2), &&\quad G(u-\omega)=G(u)+2\, m\,K(k^2), \quad m,n\in \mathbb{Z},
\label{crossing8vBetaG}
\end{align}
where $ K(k^2) $ is the Elliptic Integral of the first kind.

Notice, however, that although this R-matrix is a deformation of the AdS$ _2 $ R-matrix, the charge conjugation matrix $ \mathcal{C} $ is not similar to the one of AdS$ _2 $. The undeformed AdS$ _2 $ R-matrix (\cite{Hoare:2014kma}) is diagonal and therefore bosons and fermions are transformed into their antiparticles by crossing symmetry. In our deformation, however, the $ \mathcal{C} $ is off-diagonal and therefore bosons are transformed into fermions and vice-versa, which is highly unusual. It is important to highlight, that in the limit $ k\rightarrow \infty $ (which lead us back to AdS$ _2 $) both \eqref{CforAdS2def} and the diagonal $ \mathcal{C} $ given in \cite{Hoare:2014kma} satisfy the equations \eqref{crossing8vb}, but the diagonal one is the correct one on that case. It would be very important to try to construct the sigma model for this deformation and try to make sense of the off-diagonal $ \mathcal{C} $ we found for arbitrary $ k $. 

For the boson-boson R-matrix there is crossing for any value of $ k $ except $ k= 1 $. The  $ \mathcal{C} $ is given by \footnote{A similar solution exists for replacing $ -1\rightarrow 1 $ in the matrix element 21 of $ \mathcal{C} $}
\begin{equation}
\mathcal{C}=\begin{pmatrix}
0 & 1\\
-1 & 0
\end{pmatrix}
\label{CforAdS2defbb}
\end{equation}
and therefore also off-diagonal. The expression for $ \sigma(u\pm \omega) $ is basically the same as the boson-fermion one, just replacing the overall factor $ i $ by $ 1 $. Also, 

\begin{align}
&\eta(u+\omega)=-\eta(u)+\pi n, &&\quad \eta(u-\omega)=-\eta(u)-\pi m,\nonumber\\
&G(u+\omega)=G(u)+\, n\,K(k^2), &&\quad G(u-\omega)=G(u)+\, m\,K(k^2), \quad m,n \, \text{are odd} .
\label{crossing8vBetaGbb}
\end{align}

By comparing expressions \eqref{crossing8vBetaG} and \eqref{crossing8vBetaGbb} it is possible to see why the boson-fermion $ k=1 $ case has crossing symmetry while  the boson-boson one has not. The reason is that $ K(k^2) $ diverges for $ k=1 $, but in the boson-fermion case we still can have $ m=n=0 $ as a solution so $ K(k^2) $ does not appear, while for the boson-boson case $ m $ and $ n $ are odd, and therefore there is no way to have a finite expression when $ k=1 $.  

\section{Conclusions}

In this paper we classified regular deformations of the $S$-matrices of the integrable AdS$_{2,3}$ string sigma models. We found that the $\mathrm{AdS}_2 \times \mathrm{S}^2 \times \mathrm{T}^6$ only admits one type of deformation. But interestingly, the AdS$_3 \times \mathrm{S}^3 \times \mathrm{M}^4$ scattering matrix admits two deformations, an elliptic and a functional one. We showed that all the deformations can be made crossing symmetric. It would also be important to know how many of our deformations correspond to moduli of the sigma models and how many would go out of the string theory \cite{OhlssonSax:2018hgc}.

There are numerous interesting directions for future research. The most pressing one would be to find a deformed string sigma model which gives rise to the new elliptic deformed $S$-matrix. Since the usual way of deforming by means of the modified classical Yang-Baxter equations gives rise to quantum deformations, a more general approach might be needed, e.g. through screening charge formalism for deformed sigma models and study screening parameter limits \cite{OSP_Sigma_2020}. A first hint at what needs to be done is the fact that half of the symmetry seems to be broken, as it is only compatible with two instead of four supersymmetry generators. It would also be interesting to carry out the Bethe Ansatz for this model.

This paper also sheds some light on the three parameter deformation \cite{Delduc:2018xug,Bocconcello:2020qkt}. In particular, we find that the quantum deformed model can only be further deformed functionally. Hence, it would be good if one could define a meaningful classical limit so that these models can be compared and gain some insight in the existence and form of the full quantum scattering matrix of the three parameter model. It is also very interesting to look if with the present approach one can identify 4-parametric deformation of $ AdS_{3}\times S^{3}\times \mathrm{M}^{4} $, which could potentially follow from higher vertex models.

In this work we have also classified the possible integrable deformations of the AdS$_{2}$ model by considering the deformations of the massive modes. We were able to write a charge conjugation matrix $\mathcal{C}  $ satisfying the crossing relations, but it still necessary to understand better such $ \mathcal{C} $, since it is of off-diagonal form. We also found a potential new quantum algebra that underlies this scattering matrix and it would be interesting to figure out its exact structure and if associated Yangian symmetries lead to known integrable deformations or create a new class.  Even more interesting would be possible extensions of our deformed scattering matrix to massless modes.

\paragraph{Acknowledgments}

\
We would like to thank B. Hoare, S. Lukyanov, M. Alfimov, F. Seibold, C. Paletta, A. Torrielli, V. Korepin, S. J. van Tongeren and A. Sfondrini for useful discussions. We would like to thank B. Hoare for several interesting discussions about the AdS$ _2 $ deformation and for pointing out a problem that made possible writing the crossing symmetry relations for it.  
MdL was supported by SFI, the Royal Society and the EPSRC for funding under grants UF160578, RGF$\backslash$R1$\backslash$181011, RGF$\backslash$EA$\backslash$180167 and 18/EPSRC/3590. 
Furthermore, AP. and ALR. were supported by the grants 
 RGF$\backslash$EA$\backslash$180167 and 18/EPSRC/3590, respectively. 
P.R. was supported in part by a Nordita Visiting PhD Fellowship and by SFI and the Royal Society grant UF160578.
\

\appendix 

\section{AdS$ _{3} $ $R$-matrices}

In order to the comparison between the deformed and underformed $ R $-matrices be performed it was necessary to rewrite the $ R $-matrices for AdS$_3\times\text{S}^3\times \text{T}^4$  \cite{Sfondrini:2014via,Borsato:2014exa}  and AdS$ _3\times \text{S}^3\times \text{S}^3\times \text{S}^1 $\cite{Borsato:2015mma} in a way that they satisfy YBE for any for arbitrary $ \gamma(p) $. Also, what we call $ \gamma(p) $ here was called $ \eta(p) $ in the referred papers in order to avoid confusion with our notation.

In this appendix, we present the exact form we used in the comparisons. 

\subsection{${\rm AdS}_{3}\times {\rm S}^3\times {\rm S}^3\times {\rm S}^1$}\label{AdS3background2}

For  $ {\rm AdS}_{3}\times {\rm S}^3\times {\rm S}^3\times {\rm S}^1 $ \cite{Borsato:2015mma} the full $ R $-matrix is composed of four blocks written according to chirality of the particles
\begin{equation}
\lR^{{\rm AB}}=f^{{\rm AB}}\chi^{{\rm AB}}\begin{pmatrix}
r_1^{{\rm AB}} &    0     &    0      & r_8^{{\rm AB}}\\
0    & r_2^{{\rm AB}} & r_6^{{\rm AB}}  &    0    \\
0    & r_5^{{\rm AB}} & r_3^{{\rm AB}}  &    0    \\
r_7^{{\rm AB}} &    0     &    0      & r_4^{{\rm AB}}
\end{pmatrix}
\label{Rblocks1}
\end{equation}
where $ \lR^{{\rm AB}}\equiv \lR^{{\rm AB}}(p,q) $, $ r_i^{{\rm AB}} \equiv r_i^{{\rm AB}}(p,q)$ and $\chi^{{\rm AB}}\equiv \chi^{{\rm AB}}(p,q) $; and $ \tA$ and $\tB $ are their chiralities i.e. ${\rm A},\,{\rm B}={\rm R},\,{\rm L}  $. Below we present the explicit form of the four blocks starting with the ones with same chirality: for RR

\begin{align}
& r_1^{{\rm RR}}=1,&& r_2^{{\rm RR}}=\sqrt{\frac{x^-_{{\rm R}}(p)}{x^+_{{\rm R}}(p)}}\frac{x^+_{{\rm R}}(p)-x^+_{{\rm R}}(q)}{x^-_{{\rm R}}(p)-x^-_{{\rm R}}(q)},\nonumber\\
& r_3^{{\rm RR}}=\sqrt{\frac{x^+_{{\rm R}}(q)}{x^-_{{\rm R}}(q)}}\frac{x^-_{{\rm R}}(p)-x^-_{{\rm R}}(q)}{x^-_{{\rm R}}(p)-x^+_{{\rm R}}(q)},&& r_4^{{\rm RR}}=\sqrt{\frac{x^-_{{\rm R}}(p)}{x^+_{{\rm R}}(p)}\frac{x^+_{{\rm R}}(q)}{x^-_{{\rm R}}(q)}}\frac{x^-_{{\rm R}}(q)-x^+_{{\rm R}}(p)}{x^-_{{\rm R}}(p)-x^+_{{\rm R}}(q)},\nonumber\\
& r_5^{{\rm RR}}=\frac{x^-_{{\rm R}}(p)-x^+_{{\rm R}}(p)}{x^-_{{\rm R}}(p)-x^+_{{\rm R}}(q)}\frac{\gamma^{{\rm R}}(q)}{\gamma^{{\rm R}}(p)}, && r_6^{{\rm RR}}= \sqrt{\frac{x^-_{{\rm R}}(p)}{x^+_{{\rm R}}(p)}\frac{x^+_{{\rm R}}(q)}{x^-_{{\rm R}}(q)}}\frac{x^-_{{\rm R}}(q)-x^+_{{\rm R}}(q)}{x^-_{{\rm R}}(p)-x^+_{{\rm R}}(q)}\frac{\gamma^{{\rm R}}(p)}{\gamma^{{\rm R}}(q)},\nonumber\\
& r_7^{{\rm RR}}=0, && r_8^{{\rm RR}}=0,
\end{align}
the LL one given by
\begin{align}
& r_1^{{\rm LL}}=1,&& r_2^{{\rm LL}}=\sqrt{\frac{x^-_{{\rm L}}(p)}{x^+_{{\rm L}}(p)}}\frac{x^+_{{\rm L}}(p)-x^+_{{\rm L}}(q)}{x^-_{{\rm L}}(p)-x^-_{{\rm L}}(q)},\nonumber\\
& r_3^{{\rm LL}}=\sqrt{\frac{x^+_{{\rm L}}(q)}{x^-_{{\rm L}}(q)}}\frac{x^-_{{\rm L}}(p)-x^-_{{\rm L}}(q)}{x^-_{{\rm L}}(p)-x^+_{{\rm L}}(q)},&& r_4^{{\rm LL}}=\sqrt{\frac{x^-_{{\rm L}}(p)}{x^+_{{\rm L}}(p)}\frac{x^+_{{\rm L}}(q)}{x^-_{{\rm L}}(q)}}\frac{x^-_{{\rm L}}(q)-x^+_{{\rm L}}(p)}{x^-_{{\rm L}}(p)-x^+_{{\rm L}}(q)},\nonumber\\
& r_5^{{\rm LL}}=\sqrt{\frac{x^-_{{\rm L}}(p)}{x^+_{{\rm L}}(p)}\frac{x^+_{{\rm L}}(q)}{x^-_{{\rm L}}(q)}}\frac{x^-_{{\rm L}}(q)-x^+_{{\rm L}}(q)}{x^-_{{\rm L}}(p)-x^+_{{\rm L}}(q)}\frac{\gamma^{{\rm L}}(p)}{\gamma^{{\rm L}}(q)}, && r_6^{{\rm LL}}= \frac{x^-_{{\rm L}}(p)-x^+_{{\rm L}}(p)}{x^-_{{\rm L}}(p)-x^+_{{\rm L}}(q)}\frac{\gamma^{{\rm L}}(q)}{\gamma^{{\rm L}}(p)},\nonumber\\
& r_7^{{\rm LL}}=0, && r_8^{{\rm LL}}=0,
\end{align}
while LR is
\begin{align}
& r_1^{{\rm LR}}=\sqrt{\frac{x^-_{{\rm L}}(p)}{x^+_{{\rm L}}(p)}}\frac{1-x^+_{{\rm L}}(p)x^-_{{\rm R}}(q)}{1-x^-_{{\rm L}}(p)x^-_{{\rm R}}(q)}, && r_2^{{\rm LR}}=1,\nonumber\\
&r_3^{{\rm LR}}=\sqrt{\frac{x^-_{{\rm L}}(p)}{x^+_{{\rm L}}(p)}\frac{x^-_{{\rm R}}(q)}{x^+_{{\rm R}}(q)}}\frac{1-x^+_{{\rm L}}(p)x^+_{{\rm R}}(q)}{1-x^-_{{\rm L}}(p)x^-_{{\rm R}}(q)}, &&r_4^{{\rm LR}}=-\sqrt{\frac{x^-_{{\rm R}}(q)}{x^+_{{\rm R}}(q)}}\frac{1-x^-_{{\rm L}}(p)x^+_{{\rm R}}(q)}{1-x^-_{{\rm L}}(p)x^-_{{\rm R}}(q)},\nonumber\\
&  r_7^{{\rm LR}}=\sqrt{\frac{x^-_{{\rm L}}(p)}{x^+_{{\rm L}}(p)}}\frac{x_{{\rm R}}^+(q)-x_{{\rm R}}^-(q)}{1-x_{{\rm L}}^-(p)x_{{\rm R}}^-(q)}\frac{\gamma^{{\rm L}}(p)}{\gamma^{{\rm R}}(q)}, && r_5^{{\rm LR}}=0,\nonumber\\
&r_8^{{\rm LR}}=-\sqrt{\frac{x^-_{{\rm R}}(q)}{x^+_{{\rm R}}(q)}}\frac{x_{{\rm L}}^+(p)-x_{{\rm L}}^-(p)}{1-x_{{\rm L}}^-(p)x_{{\rm R}}^-(q)}\frac{\gamma^{{\rm R}}(q)}{\gamma^{{\rm L}}(p)}, && r_6^{{\rm LR}}=0,
\end{align}
and RL is given by
\begin{align}
& r_1^{{\rm RL}}=\sqrt{\frac{x^-_{{\rm R}}(p)}{x^+_{{\rm R}}(p)}}\frac{1-x^+_{{\rm R}}(p)x^-_{{\rm L}}(q)}{1-x^-_{{\rm R}}(p)x^-_{{\rm L}}(q)}, && r_2^{{\rm RL}}=1,\nonumber\\
&r_3^{{\rm RL}}=\sqrt{\frac{x^-_{{\rm R}}(p)}{x^+_{{\rm R}}(p)}\frac{x^-_{{\rm L}}(q)}{x^+_{{\rm L}}(q)}}\frac{1-x^+_{{\rm R}}(p)x^+_{{\rm L}}(q)}{1-x^-_{{\rm R}}(p)x^-_{{\rm L}}(q)}, &&r_4^{{\rm RL}}=-\sqrt{\frac{x^-_{{\rm L}}(q)}{x^+_{{\rm L}}(q)}}\frac{1-x^-_{{\rm R}}(p)x^+_{{\rm L}}(q)}{1-x^-_{{\rm R}}(p)x^-_{{\rm L}}(q)},\nonumber\\
&  r_7^{{\rm RL}}=\sqrt{\frac{x^-_{{\rm L}}(q)}{x^+_{{\rm L}}(q)}}\frac{x_{{\rm R}}^+(p)-x_{{\rm R}}^-(p)}{1-x_{{\rm R}}^-(p)x_{{\rm L}}^-(q)}\frac{\gamma^{{\rm L}}(q)}{\gamma^{{\rm R}}(p)}, && r_5^{{\rm RL}}=0,\nonumber\\
&r_8^{{\rm RL}}=-\sqrt{\frac{x^-_{{\rm R}}(p)}{x^+_{{\rm R}}(p)}}\frac{x_{{\rm L}}^+(q)-x_{{\rm L}}^-(q)}{1-x_{{\rm R}}^-(p)x_{{\rm L}}^-(q)}\frac{\gamma^{{\rm R}}(p)}{\gamma^{{\rm L}}(q)}, && r_6^{{\rm RL}}=0.
\end{align}
Also, the $ \chi^{{\rm AB}} $ that appear in \eqref{Rblocks1} are given by 
\begin{equation}
\chi^{{\rm LR}}(p,q)=\left(\frac{x^+_{{\rm L}}(p)}{x^-_{{\rm L}}(p)}\right)^{-1/4}\left(\frac{x^+_{{\rm R}}(q)}{x^-_{{\rm R}}(q)}\right)^{-1/4}\left(\frac{1-\frac{1}{x^-_{{\rm L}}(p)x^-_{{\rm R}}(q)}}{1-\frac{1}{x^+_{{\rm L}}(p)x^+_{{\rm R}}(q)}}\right),
\end{equation}
where $ \chi^{{\rm RL}}(p,q) $ can be obtained from $ \chi^{{\rm LR}}(p,q) $ by doing $ \{\tL \leftrightarrow \tR\} $
while 
\begin{equation}
\chi^{{\rm RR}}=1=\chi^{{\rm LL}}.
\end{equation}
Also, notice that 
\begin{equation}
\gamma^{{\rm L}}(p+\omega)=\frac{i\, \gamma^{{\rm R}}(p)}{x^+_{{\rm R}}(p)}, \quad \text{and} \quad \gamma^{{\rm R}}(p+\omega)=\frac{i\, \gamma^{{\rm L}}(p)}{x^+_{{\rm L}}(p)},
\end{equation}
and
\begin{equation}
x^{\pm}_{{\rm L}}(p+\omega)=\frac{1}{x^{\pm}_{{\rm R}}(p)}, \quad \text{and} \quad x^{\pm}_{{\rm R}}(p+\omega)=\frac{1}{x^{\pm}_{{\rm L}}(p)}.
\end{equation}
Comparing with \cite{Borsato:2015mma} notice that their $ \bar{p} $ is equal to $ \bar{p}=p+\omega $.

\subsection{${\rm AdS}_{3}\times {\rm S}^3\times {\rm T}^4$} \label{AdS3background1}

Following \cite{Sfondrini:2014via,Borsato:2014exa}, the four blocks can be written as follows
\begin{equation}
\lR^{{\rm AB}}=\zeta^{{\rm AB}}\begin{pmatrix}
r_1^{{\rm AB}} &    0     &    0      & r_8^{{\rm AB}}\\
0    & r_2^{{\rm AB}} & r_6^{{\rm AB}}  &    0    \\
0    & r_5^{{\rm AB}} & r_3^{{\rm AB}}  &    0    \\
r_7^{{\rm AB}} &    0     &    0      & r_4^{{\rm AB}}
\end{pmatrix}
\label{RblocksT4}
\end{equation}
where $ \lR^{{\rm AB}}\equiv \lR^{{\rm AB}}(p,q) $, $ r_i^{{\rm AB}} \equiv r_i^{{\rm AB}}(p,q)$ and $\zeta^{{\rm AB}}\equiv \zeta^{{\rm AB}}(p,q) $; and $ \tA$ and $\tB $ are their chiralities i.e. ${\rm A},\,{\rm B}={\rm R},\,{\rm L}  $. Below we present the explicit form of the four blocks starting with the ones with same chirality: for RR
\begin{align}
& r_1^{{\rm RR}}=-\sqrt{\frac{x^-(q)}{x^+(q)}\frac{x^+(p)}{x^-(p)}} \frac{x^-(p)-x^+(q)}{x^-(q)-x^+(p)},
&& r_2^{{\rm RR}}=\sqrt{\frac{x^-(q)}{x^+(q)}}\frac{x^+(p)-x^+(q)}{x^+(p)-x^-(q)},\nonumber\\
& r_3^{{\rm RR}}=\sqrt{\frac{x^+(p)}{x^-(p)}}\frac{x^-(p)-x^-(q)}{x^+(p)-x^-(q)},
&& r_4^{{\rm RR}}=-1,\nonumber\\
& r_5^{{\rm RR}}=-\left(\frac{x^-(q)}{x^+(q)}\frac{x^+(p)}{x^-(p)}\right)^{3/4}\frac{x^+(p)-x^-(p)}{x^+(p)-x^-(q)}\frac{\gamma(q)}{\gamma(p)},
&& r_7^{{\rm RR}}=0,\nonumber\\
& r_6^{{\rm RR}}=-\left(\frac{x^-(p)}{x^+(p)}\frac{x^+(q)}{x^-(q)}\right)^{1/4}\frac{x^+(q)-x^-(q)}{x^+(p)-x^-(q)}\frac{\gamma(p)}{\gamma(q)}, && r_8^{{\rm RR}}=0,
\end{align}
and then for LL

\begin{align}
& r_1^{{\rm LL}}=-\sqrt{\frac{x^-(q)}{x^+(q)}\frac{x^+(p)}{x^-(p)}} \frac{x^-(p)-x^+(q)}{x^-(q)-x^+(p)},
&& r_2^{{\rm LL}}=\sqrt{\frac{x^-(q)}{x^+(q)}}\frac{x^+(p)-x^+(q)}{x^+(p)-x^-(q)},\nonumber\\
& r_3^{{\rm LL}}=\sqrt{\frac{x^+(p)}{x^-(p)}}\frac{x^-(p)-x^-(q)}{x^+(p)-x^-(q)},
&& r_4^{{\rm LL}}=-1,\nonumber\\
& r_5^{{\rm LL}}=-\left(\frac{x^-(q)}{x^+(q)}\frac{x^+(p)}{x^-(p)}\right)^{1/4}\frac{x^+(q)-x^-(q)}{x^+(p)-x^-(q)}\frac{\gamma(p)}{\gamma(q)},
&& r_7^{{\rm LL}}=0,\nonumber\\
& r_6^{{\rm LL}}=-\left(\frac{x^-(q)}{x^+(q)}\frac{x^+(p)}{x^-(p)}\right)^{1/4}\frac{x^+(p)-x^-(p)}{x^+(p)-x^-(q)}\frac{\gamma(q)}{\gamma(p)}, && r_8^{{\rm LL}}=0.
\end{align}

For the blocks of opposite chirality we have for LR

\begin{align}
& r_1^{{\rm LR}}=\sqrt{\frac{x^+(p)}{x^-(p)}\frac{x^+(q)}{x^-(q)}}\frac{1-x^+(p)x^-(q)}{1-x^+(p)x^+(q)} && 
r_3^{{\rm LR}}=\sqrt{\frac{x^+(p)}{x^-(p)}},\nonumber\\
&r_2^{{\rm LR}}=\frac{x^+(p)}{x^-(p)}\sqrt{\frac{x^+(q)}{x^-(q)}}\frac{1-x^-(p)x^-(q)}{1-x^+(p)x^+(q)}, && r_4^{{\rm LR}}=-\frac{x^+(p)}{x^-(p)}\frac{1-x^+(q)x^-(p)}{1-x^+(p)x^+(q)},\nonumber\\
& r_7^{{\rm LR}}=\left(\frac{x^+(p)}{x^-(p)}\frac{x^+(q)}{x^-(q)}\right)^{3/4}\frac{x^+(q)-x^-(q)}{1-x^+(p)x^+(q)}\frac{\gamma(p)}{\gamma(q)}, && r_5^{{\rm LR}}=0,\nonumber\\
& r_8^{{\rm LR}}=-\left(\frac{x^-(q)}{x^+(q)}\right)^{1/4}\left(\frac{x^+(p)}{x^-(p)}\right)^{3/4}\frac{x^+(p)-x^-(p)}{1-x^+(p)x^+(q)}\frac{\gamma(q)}{\gamma(p)}, && r_6^{{\rm LR}}=0.
\end{align}
and then for RL

\begin{align}
& r_1^{{\rm RL}}=\sqrt{\frac{x^-(p)}{x^+(p)}\frac{x^-(q)}{x^+(q)}}\frac{1-x^+(p)x^-(q)}{1-x^-(p)x^-(q)} && 
r_3^{{\rm RL}}=\sqrt{\frac{x^-(q)}{x^+(q)}},\nonumber\\
&r_2^{{\rm RL}}=\frac{x^-(q)}{x^+(q)}\sqrt{\frac{x^-(p)}{x^+(p)}}\frac{1-x^+(p)x^+(q)}{1-x^-(p)x^-(q)}, && r_4^{{\rm RL}}=-\frac{x^-(q)}{x^+(q)}\frac{1-x^+(q)x^-(p)}{1-x^-(p)x^-(q)},\nonumber\\
& r_7^{{\rm RL}}=\left(\frac{x^-(q)}{x^+(q)}\right)^{3/4}\left(\frac{x^+(p)}{x^-(p)}\right)^{1/4}\frac{x^+(p)-x^-(p)}{1-x^-(p)x^-(q)}\frac{\gamma(q)}{\gamma(p)}, && r_5^{{\rm RL}}=0,\nonumber\\
& r_8^{{\rm RL}}=-\left(\frac{x^-(p)}{x^+(p)}\frac{x^-(q)}{x^+(q)}\right)^{3/4}\frac{x^+(q)-x^-(q)}{1-x^-(p)x^-(q)}\frac{\gamma(p)}{\gamma(q)}, && r_6^{{\rm RL}}=0.
\end{align}

 Notice that although this is the same $ R $-matrix from \cite{Sfondrini:2014via}, as mentioned above, we did two small modifications to it: the first was to write the blocks exclusively in terms of the Zhakowski variables by using 

\begin{equation}
e^{i\,p}=\frac{x^+(p)}{x^-(p)};
\end{equation}
while the second was to rewrite the blocks $ \mathcal{R}^{{\rm AB}} $ in a way that they satisfy YBE without the need to specify an expression for $ \gamma(p) $. Notice that by assuming 

\begin{equation}
\gamma(p)=e^{\frac{i\,p}{4}}\sqrt{\frac{i\,h}{2}\left(x^-(p)-x^+(p)\right)}
\end{equation}
we can recover the $ R $-matrices in \cite{Sfondrini:2014via}. 

\subsection{Obtaining the S-matrix of ${\rm AdS}_{3}\times {\rm S}^3\times {\rm T}^4$ from ${\rm AdS}_{3}\times {\rm S}^3\times {\rm S}^3\times {\rm S}^1$}

One can actually obtain the $ R $-matrix for ${\rm AdS}_{3}\times {\rm S}^3\times {\rm T}^4$ starting from the one of  ${\rm AdS}_{3}\times {\rm S}^3\times {\rm S}^3\times {\rm S}^1$. In order to do that we need the following changes for $ x^+_{R,L}  $ and $ x^-_{R,L} $

\begin{equation}
x^-_{{\rm L}}\rightarrow x^-, \quad x^-_{{\rm R}}\rightarrow x^-,\quad x^+_{{\rm L}}\rightarrow x^+ \quad\text{and}\quad   x^-_{{\rm R}}\rightarrow x^+,
\end{equation}
for $ \gamma^{\tR,\tL} $

\begin{equation}
\gamma^{\tL}(p)=a \left(\frac{x^+(p)}{x^-(p)}\right)^{1/4}\gamma(p) \quad \text{and} \quad \gamma^{\tR}(p)=a \left(\frac{x^-(p)}{x^+(p)}\right)^{1/4}\gamma(p),
\end{equation}
with $ a $ being a constant, while $ f^{{\rm AB}} $ (in \eqref{Rblocks1}) are given by
\begin{align}
& f^{{\rm LL}}(p,q)=-\sqrt{\frac{x^+(p)}{x^-(p)}\frac{x^-(q)}{x^+(q)}}\frac{x^-(p)-x^+(q)}{x^-(q)-x^+(p)}\zeta^{{\rm LL}}(p,q),\nonumber\\
& f^{{\rm RR}}(p,q) =-\sqrt{\frac{x^+(p)}{x^-(p)}\frac{x^-(q)}{x^+(q)}}\frac{x^-(p)-x^+(q)}{x^-(q)-x^+(p)}\zeta^{{\rm RR}}(p,q),\nonumber\\
& f^{{\rm LR}}(p,q) =\frac{x^+(p)}{x^-(p)}\sqrt{\frac{x^+(q)}{x^-(q)}}\frac{1-x^-(p)x^-(q)}{1-x^+(p)x^+(q)}\frac{\zeta^{{\rm LR}}(p,q)}{\chi^{{\rm LR}}(p,q)},\nonumber\\
& f^{{\rm RL}}(p,q)=\sqrt{\frac{x^-(q)}{x^+(q)}}\frac{\zeta^{{\rm RL}}(p,q)}{\chi^{{\rm RL}}(p,q)}.
\end{align}

\bibliographystyle{nb}
\bibliography{References}

\end{document}